\documentclass[12pt,preprint]{aastex}         
\usepackage{psfig}
\def\be{\begin{equation}}
\def\ee{\end{equation}}

\def\lambar{\lambda\llap {--}}

\def\lsim{\lower 2pt \hbox{$\, \buildrel {\scriptstyle <}\over
         {\scriptstyle \sim}\,$}}
\newcommand\gsim{\buildrel > \over \sim}
\begin{document}
\newcommand{\figureout}[3]{\psfig{figure=#1,width=5.5in,angle=#2} 
   \figcaption{#3} }

\title{High-Energy Emission From Millisecond Pulsars}

\author{Alice K. Harding\altaffilmark{1},
Vladimir V. Usov\altaffilmark{2} \& Alex G. Muslimov\altaffilmark{1,3}}   

\altaffiltext{1}{Laboratory of High Energy Astrophysics, 
NASA/Goddard Space Flight Center, Greenbelt, MD 20771}
 
\altaffiltext{2}{Weizmann Institute of Science, Rehovot 76100, Israel}

\altaffiltext{3}{Current address: ManTech International Corporation, 
Lexington Park, MD 20653}


\begin{abstract}
The X-ray and $\gamma$-ray spectrum of rotation-powered millisecond pulsars is
investigated in a model for acceleration and pair cascades on open field lines
above the polar caps.  Although these pulsars have low surface magnetic fields, 
their short periods allow them to have large magnetospheric potential drops, but
the majority do not produce sufficient pairs to completely screen the accelerating 
electric field.  
In these sources, the primary and secondary electrons continue to accelerate to high
altitude and their Lorentz factors are limited by curvature and synchrotron 
radiation reaction.  The accelerating particles maintain high Lorentz factors and 
undergo cyclotron resonant absorption of radio emission, that produces and maintains 
a large pitch angle, resulting in a strong synchrotron component.  The resulting spectra 
consist of several distinct components:
curvature radiation from primary electrons dominating from 1 - 100 GeV, synchrotron radiation 
from primary and secondary electrons dominating up to about 100 MeV, 
and much weaker inverse-Compton radiation
from primary electrons at 0.1 - 1 TeV.  We find that the relative size of these 
components depends on pulsar period, period derivative, and neutron star mass and radius with 
the level of the synchrotron
component also depending sensitively on the radio emission properties.  
This model is successful in describing the observed X-ray and $\gamma$-ray 
spectrum of PSR J0218+4232 as synchrotron radiation, peaking around 100 MeV 
and extending up to a turnover around several GeV.  The predicted curvature
radiation components from a number of millisecond pulsars, as well as the collective
emission from the millisecond pulsars in globular clusters, should be detectable with 
AGILE and GLAST.  We also discuss a hidden population of X-ray-quiet and radio-quiet 
millisecond
pulsars which have evolved below the pair death line, some of which may be detectable
by telescopes sensitive above 1 GeV.
\end{abstract} 


\pagebreak
  
\section{INTRODUCTION}

A relatively large fraction of the rotation-powered millisecond radio pulsars have been detected
at X-ray energies.  Of the roughly 60 pulsars with detected X-ray emission, more than half have 
periods less than 10 ms (Becker \& Aschenbach 2002).  Pulsations have been seen in about a dozen 
millisecond pulsars (MSPs).  The majority of these sources have relatively hard, power-law 
spectra at 0.1 - 30 keV, which must break or turn over before about 1-100 MeV since only one, PSR J0218+4232, 
has been detected as a $\gamma$-ray pulsar (Kuiper et al. 2000).  The spectrum of PSR J0218+4232 
measured by EGRET is much softer, with photon index 2.6, than the typical $\gamma$-ray pulsar with photon
index 1.5-2.0, and it is not detected above 1 GeV.  Although it is 30 times closer, the very nearby
MSP, PSR J0437--4715, has not been detected in $\gamma$-rays though its predicted flux is well
above EGRET sensitivity, if a standard $\gamma$-ray pulsar spectrum is assumed.  
It is evident that the high-energy properties
of MSPs are quite different from those of normal pulsars.  Furthermore the high-energy properties
of field MSPs in the galactic plane seem to be quite different from the MSP in some globular
clusters.  All 16 of the known radio pulsars in the globular cluster 47 Tuc
have been detected as X-ray point sources by the Chandra Observatory (Grindlay et al. 2001),
but pulsations were not seen due to the very small number of counts from each source.  
However, it has been determined that the spectra of the MSPs in 47 Tuc are very soft and probably
thermal. Thus, there are a number of mysteries concerning why MSPs with rotational energy-loss
very similar to those of younger high-energy pulsars behave so differently.

In this paper we attempt to address these differences by studying the electrodynamics and
emission processes of MSPs in a polar cap (PC) model.  The model is based on the 
acceleration and pair creation above pulsar PCs investigated by Harding \& Muslimov (1998; 2001 [HM01];
2002 [HM02]) and Harding, Muslimov \& Zhang (2002 [HMZ02]).  HM02 and HMZ02 studied the
acceleration of electrons above the neutron star (NS) surface in a space-charge limited flow model,
one-photon pair creation by curvature radiation (CR) and inverse Compton scattering (ICS)
photons, and the screening of the electric field by returning positrons. They found that
while many young pulsars produce sufficient electron-positron pairs to screen the accelerating 
electric field, thereby limiting acceleration to low altitudes over nearly the entire PC
(except in a narrow slot gap [Muslimov \& Harding 2003]), many older pulsars and nearly all MSPs 
are marginally producing pairs.  Low pair multiplicity results from either long periods,  
small surface magnetic fields or a combination of both.  The accelerating fields of MSPs are therefore 
unscreened over most or all of the PC.  One fundamental difference then in the
electrodynamics is that particles of MSPs continue accelerating to
high altitude, while particles of normal young pulsars stop their acceleration at low altitudes.
A second important difference is the much higher energy of the pair production attenuation 
spectral cutoff, due to the lower magnetic fields.
We will investigate the nature of the high-energy spectrum in the MSPs, to understand how
this difference in acceleration and attenuation properties affects the emission properties, and most 
importantly, whether it may account for the apparent observed emission characteristics.  Resonant
cyclotron absorption of radio emission, a mechanism recently proposed by Lyubarski \& Petrova (1998) 
to account for high-energy emission from fast pulsars, proves to be very efficient in production of
an X-ray to $\gamma$-ray component in the spectrum of some MSPs at high altitudes in the magnetosphere.

The unusual spectra of MSPs have been discussed by Usov (1983), who used the vacuum polar gap model 
of Ruderman \& Sutherland (1975) and noted that the curvature radiation power of primary accelerated electrons
would peak around 10 GeV.  He also outlined the characteristics of the synchrotron component of
CR pairs, but assumed that they were not accelerated.  Sturner \& Dermer (1994) modeled emission
from MSPs in a PC model, assuming that the primary electrons radiation is predominantly inverse Compton scattering. 
Zhang \&  Cheng (2003) computed spectra of MSPs in the outer gap
model, assuming that higher-order multipoles exist near the NS surface.
      
In \S 2 we discuss the accelerating electric field, pair production and screening in MSPs.  
We present the expected radiation characteristics of the curvature emission from accelerating
primary electrons and of the inverse Compton scattering of primary electrons on thermal
X-rays from a hot PC in \S 3 and \S 4.  In \S 5, we explore the general properties
of the electron-positron pairs created by the primary radiation, and of the synchrotron
radiation from the accelerating primary and secondary particles.  Numerical cascade simulations of these spectral components
and models for several of the known X-ray MSPs are presented in \S 6.  In \S 7 and \S 8 we give
predictions for future detections of 1-10 GeV $\gamma$-ray emission from MSPs in globular
clusters and radio-quiet MSPs below the radio/pair death-line.

\section{Acceleration of particles above the PC in MSPs}

The accelerating electric field at height $\eta - 1 \gsim \theta_0$ (scaled by the stellar radius) 
can be written as (HM98, equation [14])

\be
E_{\parallel } = 7\times 10^5~ {{B_{8,0}}\over {P_{\rm ms}^2}} \kappa _{0.15}~b,
\label{Epar-1}
\ee
where
\be
b=\left[{\cos \chi\over \eta^4} + {1\over 4} {\theta(\eta) \over \kappa }H(\eta) \delta (\eta) 
\xi \sin \chi \cos \phi_{\rm pc} \right](1-\xi ^2).
\label{b}
\ee

\noindent
Here $B_{8,0} = B_0/10^8$ G, $B_0$ is the surface value of the NS magnetic field strength; $P_{\rm ms} = P/1~{\rm ms}$ 
is the pulsar spin period in ms; $\kappa \approx 0.15 I_{45}/R_6^3$ is the general relativistic factor 
originating from the effect of inertial frame dragging, where $I_{45} = I/10^{45}\,\rm g\, cm^{2}$ is the
NS moment of inertia and $R_6 = R/10^6$ cm is the NS radius; $\eta = r/R$ is the radial distance 
in units of stellar radius,
$R$; $\xi = \theta /\theta _0$ is the dimensionless latitudinal coordinate; $\theta _0 = 
[\Omega R/cf(1)]^{1/2}$ is the half-angle of the PC, ($f(1) \sim 1 + 3r_g/4R$, when $r_g$ is the gravitational 
radius), $\Omega $ is the pulsar angular velocity; $\chi $ is the pulsar obliquity; $\phi_{\rm pc}$ is the magnetic azimuthal angle; and $H(1)$ and $\delta (1)$  are general-relativistic correction factors of order 1. 

Using the above electric field and assuming dipole field geometry, HMZ02 computed death lines in 
the $P$-$\dot P$ diagram for pair creation above pulsar PCs, shown in Figure 1.  They found that virtually all known radio 
pulsars are capable of producing electron-positron pairs in the magnetic field near the PC by attenuation of 
either curvature radiation (CR) or inverse-Compton scattering (ICS) radiation.  The death line for producing pairs by CR lies well above that for ICS, such that most known MSPs cannot produce
pairs through CR because
their surface magnetic fields are too low.  However, since ICS radiation from the accelerated primary electrons 
produces photons of much higher energy than CR, these MSPs are capable of producing ICS pairs.  HM01 and HM02 
found that CR pairs (at least for pulsars above their CR-pair death line) are produced 
with high enough multiplicities to screen the accelerating electric field, thus turning off particle acceleration
above a pair formation front, but that ICS pairs do not have high enough multiplicity to completely screen
the accelerating field.  {\bf The important result is that for pulsars below the CR-pair death line, which include most MSPs, primary particles and pairs may continue to accelerate and radiate to high altitude above the PCs.}  

The absence of screening makes these pulsars more efficient in terms of being able to convert more of their
spin-down energy loss into high-energy radiation, and predicts very different spectral properties from that
of young X-ray and $\gamma$-ray pulsars, all of which lie above the CR-pair death line.  
The Lorentz factors of the particles that continue to accelerate while radiating will become radiation-reaction
limited when their energy gain from acceleration is equal to their energy loss from radiation.  This leads to
increased efficiency, because the particles are constantly being re-supplied with the energy they lose, and therefore a much harder spectrum results.  Below, we will describe the various components expected in the spectrum of MSPs: CR and ICS from primary electrons and synchrotron radiation from primary and secondary electrons.  

\section{Curvature Radiation from Primary Electrons}
  
As was noted by HMZ02, Bulik et al. (2000) and Luo et al. (2000), the acceleration 
of primaries in MSPs can be limited by curvature radiation reaction, when the gain in primary energy 
is compensated by the CR losses, i.e. 
\be 
e|E_{\parallel }| \sim {{2e^2\gamma ^4}\over {3\rho _c^2}},
\label{CRR}
\ee
where $\gamma $ is the primary Lorentz factor, $\rho _c=4R(c/\Omega R)^{1/2}\eta ^{1/2}/3\xi $ is the radius of 
curvature of the magnetic field line.  We take the following simplified approximation for $E_{\parallel}$, 
given by equations 
(\ref{Epar-1}) and (\ref{b}) in the aligned ($\chi = 0$) and orthogonal ($\chi = \pi/2$) cases,
\be
E_{\parallel } \simeq
\left\{ \begin{array}{ll}
5\times 10^5\,B_{8,0}\, P_{\rm ms}^{-2}\, \kappa _{0.15}\,\eta^{-4}, & \chi = 0 \\ 
2.5 \times 10^5\, B_{8,0}\, P_{\rm ms}^{-5/2}\, \eta^{-1/2}, & \chi = {\pi\over 2} 
\end{array}
\right. \label{Epar-2}
\ee
Muslimov \& Harding (2004) have formally extended the solution for $E_{\parallel}$ in
the open field region to very high altitudes, up to $0.5 - 0.7\eta_{LC}$.  They find that for MSPs
the form given in equation (\ref{Epar-2}) is fairly accurate even up to these high altitudes.

From equation (\ref{CRR}), we can estimate the Lorentz factor corresponding to the regime of CR-reaction limited 
acceleration, 
\be 
\gamma _{_{\rm CRR}} \approx 10^7\, B_{8,0}^{1/4} 
\left\{ \begin{array}{ll}
\kappa _{0.15}^{1/4}\,P_{\rm ms}^{-1/4}\,\eta^{-3/4} & \chi = 0 \\
\,P_{\rm ms}^{-3/8}\,\eta^{1/8}, & \chi = {\pi\over 2}
\end{array}
\right.
\label{gammaCRR}
\ee  
The CR spectrum is hard, with photon index $-2/3$, up to a cutoff energy determined by the 
characteristic energy of the spectrum, $\varepsilon _{\rm cr} = {{3 \lambda_c\gamma _{_{\rm CRR}}^3}/{2\rho_c}}$ (where $\lambda _c \equiv \hbar/(mc)=2.4\times 10^{-10}$ cm is the electron Compton wavelength) which
(in units of $mc^2$) is 
\be
\varepsilon _{\rm cr} \approx 
(1.4 - 2) \times 10^4\, \,B_{8,0}^{3/4}
\left\{ \begin{array}{ll}
\kappa _{0.15}^{3/4} P_{\rm ms}^{-5/4}\, \eta^{-11/4}, & \chi = 0 \\
P_{\rm ms}^{-13/8}\, \eta^{-1/8}, & \chi = {\pi\over 2}
\end{array}
\right.
\label{ecr}
\ee
The $\nu F_{\nu }$ spectrum will have index $4/3$ and peak at $\varepsilon _{\rm peak} 
= 4~\varepsilon _{\rm cr}/3$ 

The instantaneous CR photon spectrum of the primary electrons is
\be
N_{CR}(\varepsilon) = {\alpha \over (\lambar mc)^{1/3}} \left({c\over \rho_c}\right)^{2/3}\,\varepsilon^{-2/3},
~~~~\varepsilon < \varepsilon_{\rm cr},
\label{insCR}
\ee
where $\lambar = \lambda_c/2\pi$.
Note that the CR spectrum does not depend on the electron energy, which only determines the cutoff
energy.
The total CR spectrum of an electron as it radiates along a dipole field line is
\be
N^{tot}_{CR}(\varepsilon) = {1\over c}\, \int_1^{\eta_{\rm max}} N_{CR}(\varepsilon) d\eta, 
~~~~\varepsilon < \varepsilon_{\rm cr}.
\label{CRint}
\ee
The upper limit on the integral, $\eta_{\rm max}$, will depend on whether the accelerating electric field is
screened or unscreened.  In the unscreened case, as we will show in \S 5, $\eta_{\rm max}$ will be set by the 
altitude at which the resonant absorption begins.  At that point, the pitch angle of the electrons increases
rapidly and $\gamma$ decreases as the electron loses energy to synchrotron emission.  As $\gamma$ decreases,
the CR losses quickly become insignificant. 
Performing the integral in equation (\ref{CRint}), the total CR spectrum of a primary electron is then,
\be
N^{tot}_{CR}(\varepsilon) = 2.5 \times 10^3\,P_{\rm ms}^{-1/3}\,E_{\rm MeV}^{-2/3}\,\eta_{\rm max}^{2/3},
~~~~\varepsilon < \varepsilon_{\rm cr}
\label{NCRtot}
\ee 
where $E_{\rm MeV}$ is the photon energy in MeV.  The total spectral flux from the PC is
$F_{CR}(\varepsilon)= N^{tot}_{CR}(\varepsilon)\,\dot N_p / \Omega d^2$,
where 
\be
\dot N_p = 1.3 \times 10^{32}\,P_{\rm ms}^{-2}\,B_{8,0}\,\rm s^{-1}
\label{np}
\ee
is the Goldreich-Julian current of
primary particles from the PC.  The flux of CR from each PC is then
\be
F_{CR}(\varepsilon)\simeq 4 \times 10^{-8}\,B_{8,0}\,P_{\rm ms}^{-7/3}
\,\eta_{\rm max}^{2/3}\,\Omega_{sr}^{-1}\,d_{\rm kpc}^{-2}\,E_{\rm MeV}^{-2/3}~~~ \,{\rm ph\, cm^{-2}\, s^{-1}\, MeV^{-1}}~~~~~~~\varepsilon < \varepsilon_{\rm cr},
\label{FCR}
\ee
where $\Omega_{sr}$ is the solid angle for each PC and $d_{\rm kpc}$ is the distance to the source in kpc.
Table 1 shows the predicted $\varepsilon_{\rm peak}$ and $F_{CR}$ for the millisecond X-ray pulsars,
assuming $\Omega_{sr} = 1$.  However, $\Omega_{sr} \lsim 2\pi$ for those pulsars with unscreened $E_{\parallel}$
that have emission extending to high altitude.

\section{Inverse Compton Radiation from Primary Electrons}

Accelerated primary electrons will also scatter thermal X-ray photons emitted from a hot PC near the
NS surface to form an ICS radiation component.  Such radiation from MSPs has been studied by Bulik 
et al. (2000), who noted that the ICS spectrum of thermal photons scattered by primary electrons will be
quite hard and extend to several TeV.  But even for the case of a hot PC at temperature 3 MK, the level
of this radiation is much lower than that of the CR component, and not detectable with current Cherenkov
telescopes.  However, HM02 showed that the ICS pairs are very important for heating the PC and
can produce detectable thermal X-ray emission. The pairs from ICS photons will also produce a synchrotron
radiation component. 

In order to compute the ICS radiation from primary electrons, we follow the approach of HM98 and HM02, 
who simplify the geometry of scattering thermal radiation from a hot PC of size $R_T$ by assuming 
that the electron moves upward along the magnetic axis.  Since the magnetic field in MSPs is low, resonant
scattering (e.g. Dermer 1990) will not be important, and only non-resonant ICS will contribute.
Since the Lorentz factors of the electrons are very large ($\gamma_{CRR} \sim 10^7$) the scattering occurs in the
Klein-Nishina limit.  The spectrum is extremely hard as the bulk of the scattered photons lie near the
kinematic maximum at the primary particle energy $\varepsilon_{ICS}^{\rm max} \sim \gamma_{CRR}$.  Details
of the method of calculation may be found in HM98 and HM02.  Here, we assume that the radius 
of the hot PC is 
the standard PC radius, $R_T = R\sin\theta_{PC} = R(2\pi R/Pc)^{1/2}$ .

The energy loss rate from inverse Compton emission is small compared to that from CR.  Evaluating the
RHS of equation (\ref{CRR}), the CR loss rate of the primary particle energy is 
\be
\dot \gamma_{CR} = 6 \times 10^{12}\,{\rm s^{-1}}\,\gamma_7^4 P_{ms}^{-1}
\ee
where $\gamma_7 \equiv \gamma/10^7$.  The ICS loss rate in the Klein-Nishina limit is approximately
(Blumenthal \& Gould 1970)
\be
\dot \gamma_{KN} \simeq 7.6 \times 10^8\,{\rm s^{-1}}\,T_6^2\,[8.85+\ln(\gamma_7\,T_6)],
\ee
where $T_6 \equiv T/10^6$ K, ignoring the effect of the decrease in density and angles of the soft photons
with increasing distance from the surface (thus providing an upper limit to the actual ICS loss rate).
For particle energies $\gamma \gsim 10^6$, ICS losses of the primary electrons can be neglected compared
to CR losses.  Furthermore, the ICS loss rate is always much smaller than the acceleration gain rate while the
electron accelerates to the CR reaction-limited energy $\gamma_{CRR}$.  Therefore ICS losses of the primary
electrons may be completely neglected.   

\section{Synchrotron Radiation by Primary and Secondary Particles} \label{sec:SR}

Electron-positron pairs are produced by 1-photon pair creation, well above the threshold of 
$2mc^2/\sin\theta_{kB}$ in the low fields of MSPs, when the condition 
\be
B'\varepsilon \sin\theta_{kB} \gsim 0.1,
\label{pair_cond}
\ee
is satisfied, where $B'$ is the magnetic field strength in units of the critical field $B_{\rm cr} = 4.4 \times 10^{13}$ G and
$\sin\theta_{kB}$ is the angle between the photon propagation and magnetic field directions.  Since the
photons are emitted essentially parallel to the local field, they do not satisfy the above condition until they 
have traveled a pathlength $s$ and acquire a sufficient angle $\sin\theta_{kB} \sim s/\rho_c \gsim 0.1/B'
\varepsilon$.   Well above threshold, the pairs have half of the energy and the same direction as that of the 
parent photon.  Assuming that their initial pitch angle $\sin \psi _{\pm } \sim \sin \theta _{kB}$, from equation (\ref{pair_cond}) we estimate that
\be
\sin \psi _{\pm } \sim 0.1/B'\varepsilon.
\label{psi}
\ee

MSPs can produce pairs from ICS photons, and those that are near and above the CR-pair
death line can additionally produce pairs from CR photons.  These pairs will have a spectrum of energies, with 
a minimum at roughly half of the photon escape energy 
\be 
\varepsilon_{\rm esc } \sim 10^6~P_{\rm ms}^{1/2}B_8^{-1},
\label{Eesc}
\ee
which is the minimum energy of photons that are absorbed by the 
magnetic field and converted into an electron-positron pair in the pulsar magnetosphere (see e.g. Harding 2000). 
Here $B_8=B_{8,0}\eta^{-3}$.  
Because the CR spectrum falls exponentially above the peak energy given by equation (\ref{ecr}), which is
near or below $\varepsilon_{\rm esc}$, most CR photons produce pairs at escape energy, such that the
initial pitch angles of the CR pairs will be
\be
\sin \psi^{CR}_{\pm } \sim {0.1\over B'\varepsilon_{\rm esc}} \sim 0.04\,P^{-1/2}_{\rm ms}.
\label{psi_CR}
\ee
The ISC photons will produce most pairs at the maximum of the ICS spectrum, at $\epsilon_{ICS}^{\rm max} \sim 
\gamma_{_{CRR}}$, so that the initial pitch angles of ICS pairs will be
\be
\sin \psi^{ICS}_{\pm } \sim {0.1\over B'\gamma_{_{CRR}}} \sim 0.05\,B^{-5/4}_{8,0}
\left\{ \begin{array}{ll}
\kappa_{0.15}^{-1/4}\,P_{\rm ms}^{1/4} & \chi = 0  \\
P_{\rm ms}^{3/8} & \chi = {\pi\over 2}
\end{array}
\right.
\label{psi_ICS}
\ee

For a pulsar below the CR-pair death line, the multiplicity (number of pairs per primary) of either CR
or ICS pairs is not sufficient 
to screen the accelerating electric field beyond the pair-formation front (see e.g. HM02), 
over at least a large part of the PC, so that the primary and secondary electrons
continue accelerating.  The Lorentz factors, $\gamma$, and perpendicular momentum, $p_{\perp}$ (in units of $mc$), 
of each particle will evolve along the field lines according to its equation of motion, which
may be written
\be \label{dgamma}
{d\gamma\over dt}={eE_\parallel\over mc}-{2e^4\over 3m^3c^5}
B^2\,p_\perp^2 - {{2e^2\gamma ^4}\over {3\rho _c^2}}
+ \left({d \gamma\over dt}\right)^{abs} \,
\ee

\be  \label{dp_perp}
{d p_\perp\over dt}=
-{3\over 2}{c\over r}{p_\perp}
-{2e^4\over 3m^3c^5}B^2\,{p_\perp^3\over \gamma} + \left({d p_\perp(\gamma)\over dt}\right)^{abs}.
\ee
The terms of the right hand side of equation (\ref{dgamma}) are acceleration, synchrotron losses,
curvature radiation losses and cyclotron/synchrotron absorption.  The terms of the right hand side of equation (\ref{dp_perp}) are adiabatic changes along the dipole field line, synchrotron losses and 
cyclotron/synchrotron resonant absorption.  A derivation of the above equations (minus the CR loss and 
resonant absorption terms) is given in the Appendix.  
As we showed in \S 4, ICS losses may be neglected for the primary particles and may also be neglected
for the pairs since the acceleration and synchrotron loss rates are much larger. 
We will derive here the terms for cyclotron/synchrotron resonant absorption.

The cyclotron resonant absorption of radio emission by relativistic particles in pulsar magnetospheres,
followed by spontaneous synchrotron emission, 
was first proposed some years ago by Shklovsky (1970) as a mechanism for generating the optical
radiation from the Crab pulsar.  The process involves the absorption of photons at the cyclotron 
resonant frequency in the rest frame of the particle, resulting in an increase in the particle pitch angle.
The particle then spontaneously emits cyclotron or synchrotron radiation, depending on whether its
momentum perdendicular to the magnetic field, in the frame in which the parallel momentum vanishes, is
non-relativistic or relativistic.   Blandford \& Scharlemann (1976) computed the cross section for 
cyclotron resonant absorption, but their application of the process to the Crab pulsar resulted in
too small a re-radiated cyclotron radiation flux to explain the Crab optical emission.  However, they assumed
that the perpendicular momentum remained non-relativistic, in which case the applicable rate is that of 
cyclotron emission from the first excited Landau state, which is small relative to the rate from 
highly excited states.
Shklovsky (1970) in fact had discussed the possibility of synchrotron absorption to high Landau states
and the consequent use of the much higher synchrotron emission rates to account for the Crab pulsar optical
emission.  But the mechanism was deemed to be untenable by O'Dell \& Sartori (1970), arguing that since the 
pitch angles of the particles are very small the spectrum would have a natural cutoff of $B'/\sin\gamma$,
which is above optical frequencies.  This limit on the small-pitch angle synchrotron spectrum essentially
come from the fact that the particle cannot emit a photon with energy smaller than the cyclotron frequency
in the frame of pure circular motion.  Epstein \& Petrosian (1973) added additional criticisms involving
predicted time variations and polarization characteristics which were inconsistent with the observations.
The model was laid to rest until its revival a few years ago by Lyubarski \& Petrova (1998, LP98), who performed
a more detailed analysis of the distribution functions of particles undergoing synchrotron resonant absorption 
of radio photons and found that the particles can increase their pitch angles rapidly enough in the outer magnetosphere to attain relativistic perpendicular momentum.  The resulting synchrotron radiation can 
explain the optical emission spectrum of the Crab and other young pulsars (Petrova 2003).  

As we will show, cyclotron resonant absorption of radio emission can work very efficiently for MSPs, 
especially for those pulsars where the accelerating 
electric field is unscreened.  In fact the particles can reach the resonant absorption condition
\be  \label{rescond}
B' = \gamma\varepsilon_0\,(1-\beta\mu)
\ee
much closer to the NS surface, where $\varepsilon_0$ is the energy of the radio photon in the lab frame
(in units of $mc^2$), $\beta = (1-1/\gamma^2)^{1/2}$,
$\mu =\cos \theta$, and $\theta$  is the angle in the lab frame between the photon direction 
and the particle momentum, which is to good 
approximation the same as the magnetic field direction.  In addition, we will see that a continuously
accelerating particle can stay in the resonance, finding an equilibrium between gain in pitch angle through
resonant absorption and the loss in pitch angle through synchrotron emission.  The continuous pumping
of energy into perpendicular momentum makes the mechanism extremely efficient for MSPs.  
 
In order to determine the rate of increase of particle Lorentz factor and perpendicular
momentum in equations (\ref{gamma}) and (\ref{p_perp}), we first estimate the rate of 
perpendicular momentum gain of a single electron due to cyclotron resonant absorption.
The rest-frame total cross section for cyclotron absorption from the ground
state to the first excited Landau state is (e.g. Blandford \& Scharlemann 1976, 
Daugherty \& Ventura 1978, Harding \& Daugherty 1991)
\be  \label{cycabs}
\sigma_{abs}(\varepsilon, \theta) = \alpha\,\pi^2 \lambar^2
\,\delta(\varepsilon' - B')\,(1 + \mu'^2)
\ee
where $\varepsilon'$ and $\mu' = \cos\theta'$ are the incident soft photon energy and
angle to the local magnetic field $B'$ in the particle rest frame and 
$\alpha = e^2/\hbar c = 1/137$ is the fine structure constant.
For this calculation, we can assume that $(1 + \mu'^2) \simeq
1$.  Of course, the electron will not be absorbing photons only from
the ground state if the rate of absorption is initially much higher
than the rate of spontaneous re-emission.  This will be true for low
fields in the outer magnetosphere, so that the above cross-section
would not necessarily be valid and one should use the more complicated
cross section for absorption from an arbitrary excited Landau state.
However, LP98 have found in a relatively
careful analysis of the full classical cross section that for the case
of pulsar magnetospheres, as long as $\beta_\perp < \theta$, which is true 
when the absorbed energy is small compared to the total electron energy,
then: 1) absorption at the first harmonic dominates even for transitions at
high Landau states and 2) the cross section is the same as that in equation 
(\ref{cycabs}).  This greatly simplifies the calculation.

The spectral intensity of radio photons in the lab frame is
\be  \label{nR}
I(\varepsilon,\Omega) \simeq I_0\left({\varepsilon \over \varepsilon _0} \right)^{-\nu} 
\delta(\mu-\mu_0)\delta(\phi-\phi_0),~~~~~~~~~~\varepsilon > \varepsilon_0
\ee
where $I_0$ is the normalization value for the intensity which will be defined below, 
and the delta-function for $\mu$ is a simplification that assumes the 
radio emission is generated at a large distance from the site of resonant absorption 
and much nearer the NS surface.  

According to LP98, the angle that the radio photon direction makes with the relativistic electrons 
in the resonance region is approximately
\be \label{mu0}
\theta_0 \sim {(\eta-\eta_R)\over 2 \eta_{LC}}\sin\chi \pm {3\over 4}
\left({\eta\over \eta_{LC}}\right)^{1/2},
\ee
where $\eta_{LC} = r_{LC}/R$ 
is the dimensionless light cylinder radius, $\eta _R$ is the dimensionless radial coordinate of 
the radio emission site, and $\chi$ is the pulsar
inclination angle.  The first term in equation (\ref{mu0}) is due to the
NS rotation, while the second term is due to the field line curvature.  
The NS rotation is a larger effect than the curvature of the field lines in the outer 
magnetosphere, but the second term can be important also for MSPs at smaller radii. 

Our approach differs significantly from that of LP98, in that we solve the equation of motion of
single, continuously accelerating particles, while they computed changes to the plasma distribution 
function of non-accelerating particles.  However, the rate of increase of the pitch angle due to resonant
absorption (the final term on the RHS of [\ref{dp_perp}]) is governed by diffusion 
in momentum space and requires a solution of the kinetic equation.  Petrova (2002) has derived the
solution for the distribution function of electrons undergoing resonant absorption of radio emission in a
pulsar magnetosphere and the corresponding mean-square value of the pitch angle.
According to equation (2.17) of Petrova (2002), the mean square of the pitch angle can be calculated as 
\be
\langle {\psi ^2} \rangle = 4\,R\int _{\eta _R}^{\eta } {\it a}(\eta ')d\eta ' ,
\label{psi-msq}
\ee
where 
\be
{\it a}(\eta ) = {{2\pi ^2 {\it e}^2 (1-\beta\mu_0) I_0}\over {\gamma ^2 m^2c^4}} 
\left( {{\varepsilon _0 \gamma (1-\beta\mu_0)}\over {B'}} \right) ^{\nu },~~~~  \eta > \eta_R.
\label{a}
\ee 
Here $I_0$ is the intensity of observed radio emission measured in $erg\cdot cm^{-2}\cdot s^{-1}\cdot Hz^{-1}$. 
Thus, for the perpendicular momentum change due to cyclotron resonant absorption we can write 
\be
\left({{dp_{\perp }}\over {dt}}\right)^{abs} = 2~{\it a(\eta)}~c {{\gamma ^2}\over {p_{\perp}}} + 
{{p_{\perp}}\over p} \left({{dp}\over {dt}}\right)^{abs} 
\label{dp_dt_eval}
\ee
where we used the relationship $p_{\perp } = p {\langle {\psi ^2} \rangle}^{1/2}$.  Thus, we 
assume that $p_{\perp }$ is proportional to the root mean-square value of the pitch angle.  
We also make the further approximation of computing the evolution of the root mean-square value
of $p_{\perp }$ rather than the evolution of the particle distribution function.  Since the 
primary and secondary electrons are continuously accelerating, $\gamma$ remains very high and 
$p_{\perp }/p = \sin\psi \ll 1$.  According to Petrova (2003), the width of the $p_{\perp }$ 
distribution is of order $p_{\perp }$, so that the large variations in $\gamma$ and $p$ in r 
along the field lines is much more important in the formation of the spectrum that the spread in the
$p_{\perp }$ distribution.
 
By substituting equation (\ref{dp_dt_eval}) into the right hand sides of equations 
(\ref{dgamma}) and (\ref{dp_perp}), we get 
\be  \label{gamma}
{d\gamma\over dt}=a_1 E_{\parallel,5}- b_1B_8^2\,p_\perp^2
-c_1\gamma^4 
\ee 
\be  \label{p_perp}
{dp_\perp\over dt}=-a_2 \eta^{-1}\,p_\perp - b_1 B_8^2\,p_\perp^3
{1\over \gamma} + \left({{dp_{\perp }}\over {dt}}\right)^{abs},
\ee
where $a_1 = 1.76 \times 10^{12}\,\rm s^{-1}$, $b_1 = 1.93 \times 10^7\,\rm s^{-1}$, 
$c_1 = 5.6 \times 10^{-3}\,\rm s^{-1}$, $a_2 = 4.5 \times 10^4\,\rm s^{-1}$, $E_{\parallel,5}=E_{\parallel}/(10^5\,{\rm e.s.u.})$. Since Petrova (2002) has assumed 
that $p$ and $\gamma$ are constant to compute the change in pitch 
angle due to resonant absorption, we have neglected the change in $\gamma$ due to 
absorption in equation (\ref{gamma}).
Both $E_{\parallel,5}$ and
$B_8$ are functions of $\eta$.
In the right hand side of equation (\ref{p_perp}) the term $(dp_{\perp}/dt)^{abs}$ 
(see equation [\ref{dp_dt_eval}]) translates into 
\be
\left({{dp_{\perp}}\over {dt}}\right)^{abs} = D{{\gamma^{\nu}}\over {p_{\perp }}} + 
{{p_{\perp } \gamma }\over {\gamma^2 -1}} \left({{d\gamma }\over {dt}}\right)^{abs},~~~~~~~~~\gamma < \gamma _R
\label{dp_perp_abs}
\ee
where
\be  \label{D}
D = 5.7 \times 10^{9}\,\rm s^{-1}\,\gamma_R^{-\nu}\,\left({d_{kpc}\over \eta}\right)^2\,
\Phi_0[{\rm mJy}]\,(1-\beta\mu_0),
\ee 
and we also can neglect the $({d\gamma }/{dt})^{abs}$ term in equation (\ref{gamma}) and (\ref{dp_perp_abs}) since it
was ignored in Petrova's derivation of $\langle {\psi ^2} \rangle$.
In the above expression, we have assumed that $I _0 = \Phi _0\Omega_{rad} d^2/A$, 
where $\Phi _0$ is the measured radio flux (in mJy), $d$ is the source distance, $\Omega_{rad} \sim A/r^2$ is 
the radio emission solid angle, with $A$ and $r$ the cross-sectional area and radius at the 
absorption radius. Also, from the resonant condition (see equation [\ref{rescond}]), $\gamma < \gamma _R$,  
$\gamma _R$ is defined as
\be
\gamma _R = {{B'}\over {\varepsilon _0(1-\beta \mu _0)}} 
= 2.8\times 10^5 {{B_8}\over {\varepsilon _{0, {\rm GHz}} (1-\beta\mu_0)}}. 
\label{gamma_R}
\ee
The resonant terms will switch on only when the resonant
condition is satisfied.  We show numerical solutions of the
above equations in Figure 2.  In the case of the secondary electrons 
that are created with finite pitch angles given by either equation 
(\ref{psi_CR}) or (\ref{psi_ICS}) and a spectrum of energies above escape energy, 
one observes three distinct stages in the particle dynamical evolution.  
In stage 1, synchrotron cooling is dominant and both $\gamma$ and $p_\perp$
decrease until the synchrotron losses equal acceleration gains, while the
pitch angle remains roughly constant.\footnote{
In the relativistic (very high Landau state) limit of synchrotron radiation, 
particles lose their energy while maintaining a constant pitch angle, because 
the photons are emitted close to the orbital plane in the Lorentz frame in which 
the particle momentum is perpendicular to the field direction.  The radiation
reaction force is therefore in the opposite direction to the particle momentum in 
the lab frame, to a very good approximation.}  
In stage 2
acceleration dominates and the particle gains parallel energy and momentum
while $p_\perp$ remains roughly constant, until curvature radiation balances
acceleration.  In stage 3, the $\gamma$ of the particle is limited by 
curvature-radiation reaction, as described in \S 3, at the value given by 
equation (\ref{gammaCRR}).  During stage 3, as the particle moves to higher
altitude along the field line, B decreases until the resonant condition in 
equation (\ref{gamma_R}) is satisfied.  At the beginning of stage 4, the resonant 
absorption term will turn on and increase $p_\perp$ until the synchrotron
losses equal the acceleration gain.  During stage 4, the particle $\gamma$
can remain in equilibrium between synchrotron losses and acceleration gains, while
the $p_\perp$ stays in equilibrium between synchrotron absorption and losses,
at altitudes up to the light cylinder. 
The primary electrons have essentially zero pitch angle initially, so
they start in stage 3 and from there follow a similar evolution to the
secondary electrons.  The secondary positrons follow stage 1, in the same
manner as the secondary electrons, but in stage 2, they decelerate and are 
turned around and accelerate back toward the NS surface.  Therefore, they may
get into stage 3 (although going in the opposite direction), but do not 
participate in the synchrotron absorption of stage 4.  Figure 2a shows that the condition 
$\sin\psi \ll \theta_0$ is always satisfied, so the assumption, that the absorption cross section
of equation (\ref{cycabs}) is valid throughout the particle evolution, is justified.

In the case of a power law radio spectrum, then, a steady-state is established 
between synchrotron losses and acceleration gains for $\gamma$, and between
synchrotron absorption and losses for $p_\perp$, for both secondary and 
primary electrons at high enough altitudes.  One can find the steady-state regime in
which synchrotron absorption balances synchrotron losses in equation (\ref{p_perp}) can
be approximated by ignoring the last two terms in equation (\ref{gamma}) and the 
first term in equation (\ref{p_perp}) to give, from equation (\ref{gamma}),
\be \label{p_perpSRR}
p_\perp^{_{SRR}} \simeq \left({a_1 E_{\parallel,5}\over b_1 B_8^2}\right)^{1/2} =
302\,B_8^{-1}\,E_{\parallel,5}^{1/2}
\ee
and from equation (\ref{p_perp})
\be \label{gammaSRR}
\gamma_{_{SRR}} \simeq \left({b_1 B_8^2\over D}\right)^{1/(\nu+1)} 
(p_\perp^{_{SRR}})^{3/(\nu+1)} 
\ee
The steady-state synchrotron critical energy is then
\be \label{epsSRR}
\varepsilon_{SRR} \simeq {3\over 2}\,\gamma_{_{SRR}}\,\,p_\perp^{_{SRR}}\,B' 
\ee

We find then that the particles do reach relativistic $p_\perp$ due to synchrotron resonant
absorption, in agreement with the results of LP98.   When the particles are also continuously
accelerating, they can maintain relativistic $p_\perp$ over a large range of radii up to the
light cylinder, radiating significant amounts of high-energy synchrotron emission.  
In the next section we will compute the spectrum of this radiation. 

It should be noted that our treatment of
the particle dynamical evolution is not completely self-consistent.
In calculating the rate of increase
of the pitch angle due to resonant absorption of radio
photons we use equation (\ref{psi-msq}) obtained by Petrova (2002)
under the assumption that other changes of the particle
momentum are negligible, whereas this assumption is
not justified in our case. In particular, the Lorentz factor
of particles, $\gamma$, varies significantly in the process of
their outflow (see Fig. 2a).  However, the use of equation 
(\ref{psi-msq}) does not introduce a significant error if the length
scale, $L_{p_\perp}$, for changes in $p_\perp$ are small compared to the scale,
$L_{\gamma}$, for changes in $\gamma$.  Taking the logarithmic derivative of
$\gamma_{_{SRR}}$ in equation(\ref{gammaSRR}), we get $L_{\gamma} \sim (\nu+1)/(5+3\nu)\eta$
(in stellar radius units),
so that variations in $\gamma$ occur on a scale comparable to the NS radius.
Taking the logarithmic derivative of $(p_{\perp})^{abs}$ in equation(\ref{dp_perp_abs})
we obtain $L_{p_\perp} = c(p_\perp^{SRR})^2/R\,D\gamma_{_{SRR}}^{\nu}$.
Evaluating this result for the parameters of PSR J0218+4232 and $\nu=2$, used in Fig. 2,
we have $L_{p_\perp} \sim 0.02\,\eta^{8/3}/(1-\beta\mu_0)$, so that
variations in $p_\perp$ due to resonant absorption do indeed occur on a much smaller length 
scale for most of the particle evolution.  Based on the above analysis, it appears that
neglect of the slower changes
in $\gamma$ and $p$ due to acceleration and curvature radiation losses in estimating
the rate of resonant absorption may not introduce a very large error.  
We hope to discuss the $\psi$ diffusion
of the outflowing particles in a more general case elsewhere.

\section{Numerical Calculations of MSP Spectra}

Although the CR spectrum of the primary electrons in the radiation-reaction limit can be calculated 
analytically, as we have shown above, the synchrotron radiation spectrum of the primary and 
secondary electrons cannot be determined very accurately by analytic calculations.  
This is because the level depends on the multiplicity of both 
CR and ICS pairs and also because the spectrum is sensitive to the form of $E_{\parallel}$, unlike the
case for the CR spectrum, where only the critical energy but not the spectrum itself depends on 
$E_{\parallel}$.  The functional form of $E_{\parallel}$ for general altitude and inclination 
angle is more complicated to integrate analytically.  Furthermore, we also want to evaluate the spectra 
of MSPs above the CR death line, whose $E_{\parallel}$
is screened by pairs and the spectrum is influenced by the energy loss of the particles.

In order to compute the spectra of primary and secondary particles, we use a code that is adapted
from that used in HM01 and HM02 to calculate acceleration, radiation, pair cascades and screening 
of the $E_{\parallel}$ above the PC.  The calculation follows the dynamics of a primary electron 
as it accelerates from the NS surface, losing energy to CR and ICS, and follows the pair cascades 
that are initiated by the CR and ICS high-energy photons.  The code thus computes the 
distribution in energy and altitude of pairs of each type.  The CR and ICS radiation from the
primary electron is computed by dividing the path along a field line, identified by its magnetic
colatitude at the NS surface, into steps of fixed length. The spectrum at each step is divided into 
equal logarithmic energy intervals.  A representative
photon at the average energy of each interval is followed to its pair production or escape point.
The pairs and photons are then weighted by the value
of the spectrum at that interval, as detailed in HM02 and Harding et al. (1997). 
Photons that escape are accumulated in a distribution of energy and generation number.
If the photon produces an electron and positron pair, the radiation from each member of the
pair is treated individually, accumulating the emission from the pair
production point to a maximum radius, $r_{\rm max} = R_{LC}/\sin\chi$.  The particle equations
of motion (\ref{gamma}) and (\ref{p_perp}) are integrated along the field line starting from the
pair formation front for the primary and at the pair 
production point for the pairs.  At their production points,
the pairs have high initial energies, assumed to be equal to half of the parent photon energy, 
and pitch angles determined by the angle of the parent photon to the local magnetic field at the
pair production point.  Initially, the synchrotron radiation energy loss rate is much greater than
the gain rate due to acceleration, so the particle loses its initial energy over a relatively short
path length, $c \gamma /\dot \gamma_{SR}$.  If the MSP is below the CR pair death line, so that the 
$E_{\parallel}$ is unscreened, the decrease in $\gamma$ will stop when $\dot \gamma_{SR}$ becomes
comparable to the acceleration gain rate $e E_{\parallel}/mc$.  The particle will then begin
accelerating (electrons upward and positrons downward) and the Lorentz factor will eventually
become limited by CR reaction.  If at any time in its evolution, $\gamma > \gamma_R$ as defined 
just after equation (\ref{dp_dt_eval}) (i.e. the resonant condition is satisfied), and the particle
has reached a radius greater than the radio emission radius $\eta_R$,
then synchrotron absorption will begin contributing to the right-hand side of equation 
(\ref{p_perp}) for the $\gamma$ and $p_\perp$ evolution.  At this point,
the pitch angle begins increasing and the particle Lorentz factor drops due to synchrotron losses,
so that CR quickly becomes unimportant relative to SR.   Formation of the CR component therefore
stops at a short distance above the altitude where cyclotron absorption begins.
When the particle becomes synchrotron radiation-reaction limited, its energy 
will change more slowly with path length, governed by the change in the magnetic field and $E_{\parallel}$ 
with $\eta$.  In this case significant emission occurs at all altitudes above the NS surface. 
If the MSP is above the CR pair death line, so that the $E_{\parallel}$ is screened,
the particle continues to lose energy at the synchrotron loss rate, and most of the radiation is
emitted close to the NS surface.  The length of each step along the particle trajectory is set
dynamically so that $\gamma$ or $p_\perp$ change by a fixed fraction $f$ of their values, whichever 
gives the smaller $ds$ as determined by $ds = fc\,d\gamma/\gamma $ or $ds = fc\, dp_\perp/p_\perp$. 
The primary electrons will also undergo significant cyclotron absorption when the conditions
$\gamma > \gamma_R$ and $\eta > \eta_R$ are satisfied.

At each step, the particle radiates an instantaneous synchrotron spectrum,
given by (Tademaru 1973)
\be
\dot n_{SR} (\varepsilon) = {2^{2/3}\over \Gamma({1\over 3})}\,\alpha B' \sin\psi\, \varepsilon^{-2/3}\, \varepsilon_{_{SR}}^{-1/3}, ~~~~~~~~~~~~~~~\varepsilon < \varepsilon_{_{SR}}.
\label{nSR}
\ee
where $\sin\psi = p_\perp/p$ and $p^2 = \gamma^2 - 1$.
The spectrum at each step is divided into equal logarithmic energy intervals and sample photons are 
traced to pair production or escape points, as in the case of the primary CR and ICS spectra. 
The synchrotron radiation from each generation of pairs is computed as described above, with the 
weights from each generation multiplying the accumulated weights from previous generations.

The resulting model high-energy spectrum for the case of the radio, X-ray and $\gamma$-ray pulsar 
PSR J0218+4232 is shown in Figure 3.  The multiple components of the spectrum, CR and ICS from 
a single primary electron and synchrotron radiation from primary and secondary electrons, 
are displayed separately in Figure 3a.  As expected from the analytic results of Section \ref{sec:SR}, 
the unattenuated total CR spectrum peaks around 10 GeV, although the attenuated CR spectrum peaks around 
5 GeV (as shown),
and the ICS spectrum is quite hard, peaking at the maximum kinematic energy at nearly 10 TeV.
However, an observer who detects the synchrotron component at higher
altitude does not see the CR from lower altitudes which is radiated at smaller angles to the 
magnetic pole.  We therefore show the CR spectrum that is emitted above the radius $\eta_R = 2$,
which is significantly lower than the total emission. 
Since this pulsar is below the CR pair death line (see Fig. 1), there are
not enough pairs to screen the $E_{\parallel}$ at all colatitudes $\xi$.  Our calculations show that
the $E_{\parallel}$ is screened for $\xi \lsim 0.7$ but unscreened at the higher $\xi$ values, due to the
decrease in $E_{\parallel}$ in the outer parts of the PC.  Over the unscreened part of the PC 
acceleration of both primary and secondary electrons continues to high altitude, where resonant cyclotron 
absorption can operate.  In PSR J0218+4232 resonant absorption is very effective, due to the unscreened
but still high electric field and its the relatively high radio luminosity.  
We chose the parameters of the radio spectrum, $\varepsilon_{GeV} = 0.4$, 
$\nu = 2.0$, $\Phi_R[{\rm mJy}] = 700$ at 400 MHz and the radio emission altitude, $\eta_R = 2.0$ to best 
fit the observed X-ray and $\gamma$-ray spectrum.  Although the observed value of radio flux at 
400 MHz is only 35 mJy, 
it is possible that there is a significant absorption along the line of sight so that the 
specific intensity in the magnetosphere is higher.  This pulsar also exhibits giant pulse emission
(Joshi et al 2004) which would produce an effectively higher (but variable) flux for cyclotron absorption.
In fact the SR component may exhibit variability as a result.  The
multiplicity of CR pairs ($\sim 0.1$ for the case shown) is higher than for ICS pairs ($\sim 10^{-4}$), 
but both are
low enough that the radiation from secondaries make a minor contribution compared to that from the primary 
electrons.  As was discussed in Section \ref{sec:SR}, the number of CR pairs depends on the position
of the peak of the CR spectrum relative to the pair escape energy.  Even though the CR photons have
much lower energies than the ICS photons, the number of CR photons is much larger, so that CR
pairs dominate if $\varepsilon_{\rm CR} \lsim \varepsilon_{\rm esc}$.  
The peak of the synchrotron spectra though are 
determined by the critical SR energy of the particles when they initially reach the steady state given 
approximately by eqs (\ref{p_perpSRR})-(\ref{epsSRR}).  The SR spectrum also depends on the pulsar
inclination angle $\chi$ and magnetic azimuth $\phi_{\rm pc}$, as these parameters determine how $E_{\parallel}$
varies with altitude.  The spectra shown in Figures 3a and 3b assume $\chi = 1.0$ and $\phi_{\rm pc} = \pi/2$,
where $\phi_{\rm pc} = 0$ defines the direction toward the rotation axis.

Figure 3b shows the same model total SR spectrum and the CR spectrum emitted above $\eta_R$ 
for the parameters of PSR J0218+4232 plotted with the data.  
The model single-primary spectra are normalized to the observed flux by the factor
$\dot n_p / \Omega d^2 = 4 \times 10^{-12}\,\rm cm^{-2}\,s^{-1}$, where $\dot n_p$ is the total 
flux of primary electrons from the PC, as given by equation (\ref{np}), $d = 5.85$ kpc is the distance to the 
source and a solid angle of $0.2$ sr is assumed.  This model thus can account for the flux and spectrum in the X-ray bands but slightly violates the EGRET upper limit above 1 GeV.  The low-energy
part of the SR spectra have index 4/3, reflecting the single particle emissivity, but the high-energy
part of the SR spectrum is extended into a soft power law before turning over around 100
MeV.  The shoulder on this turnover which extends beyond 1 GeV is caused by the decrease of $\gamma$
to steady-state after the initial increase in the pitch angle.
The SR spectrum, as noted in Section \ref{sec:SR}, is the combined emission from the whole range of
altitudes from near the NS surface, forming the higher-energy portion, to near the light cylinder,
forming the lower-energy portion.  For the value of inclination shown,
the critical SR energy decreases with altitude, so that the high-energy part of the SR spectrum
visible to EGRET comes from near the NS surface and the low-energy part of the spectrum visible
to X-ray detectors comes from near the light cylinder.  The peak around 100 MeV is indicative of the 
critical SR energy at the radiation-reaction limited energy.  
Because of its location in the $P$-$\dot P$ diagram and its high radio luminosity, PSR J0218+4232
produces a significant SR component which extends up to the $\gamma$-ray range.

Another MSP, PSR J0437--4715, with period P = 5.75 ms and field comparable to that of J0218+4232, 
is further below the CR death line.  This pulsar produces many fewer CR pairs because the peak of
the CR spectrum is below 10 GeV and well below the pair escape energy, as shown in Figure 4.  The
$E_{\parallel}$ in this pulsar is unscreened at all colatitudes over the entire PC.  But since the
value of $E_{\parallel}$ is much lower than that of PSR J0218+4232, due to its longer period, and
lower radio luminosity, the synchrotron emission resulting from resonant cyclotron absorption is much
weaker.  Due to the small multiplicity of ICS pairs, the synchrotron emission from primary electrons 
dominates.  Even though J0437--4715 is much closer, at a distance of only 150 pc, than J0218+4232,
we predict that the SR flux will be undetectable.  Indeed, the X-ray spectrum of this pulsar seems 
to be dominated by at least one thermal component.  The possible weak power-law component may in fact 
be an additional thermal component (Zavlin et al. 2002) and some PC heating from ICS pairs is expected 
for this pulsar (HM02).  

Figure 5 shows a model spectrum for PSR B1821--24, a 3 ms pulsar with a relatively high magnetic field
which is one of the few MSP that lie above the CR death line.  The $E_{\parallel}$ 
will be screened above the CR pair formation front, so that primary electrons and pairs will not be accelerated 
to high altitudes.  Resonant cyclotron absorption will therefore not be effective as the particles must
keep accelerating to stay in the resonance.  According to this model then, both the primary and pair spectra 
will be softened by radiation losses.  
The SR spectrum is dominated by the pairs from CR photons and has photon index -3/2, 
the expected spectrum of particles losing energy only through SR losses at constant
pitch angle (Tademaru 1973), with a high
energy turnover at around 1 GeV.  A low energy turnover at $\varepsilon \sim 10^{-4}$ occurs at the local
cyclotron energy.  Because of CR energy losses, the CR spectrum of the primary electrons is softened and depressed below the radiation-reaction limited CR spectrum, and near the EGRET upper limits.  Since
the electric field is screened close to the NS surface, most of the radiation will be emitted at
low altitude, making the solid angle relatively small compared to the MSPs with unscreened 
fields.  Indeed, a solid angle of 0.2 sr. is required to fit the observed X-ray emission level.

\section{Emission From MSPs in Globular Clusters}

A large fraction of radio MSPs are members of globular clusters.  Some 16 millisecond radio pulsars have been 
discovered in 47 Tuc, as well as several each in M6, M28 and NGC 6397, including B1821-24 discussed above.
Chandra observations of 47 Tuc have detected all of the known radio pulsars in 47 Tuc as X-ray
point sources, but not yet as pulsed sources.  The X-ray spectra are significantly softer than the
spectra of MSPs in the field or in other clusters, and the X-ray luminosity, $L_x$, has a different
dependence on spin-down energy, $\dot E_{SD}$ (Grindlay et al. 2001).  The fact that the 47 Tuc pulsars
are older and lie further below the CR pair death line than the field pulsars, J0218 and B1937, that have 
non-thermal X-ray spectra may explain the dominance of thermal emission (possible from PC
heating) in their spectra.  Their spectral properties are closer to those of the field pulsars
J0437--4715 and J2124-3358 that also lie well below the CR death line.

High-energy $\gamma$-ray telescopes may detect the collective emission from the MSPs in globular
clusters, even though they are too distant to be detected individually.  In fact such telescopes
do not have the required angular resolution ($\sim 1''$) to distinguish the individual pulsars in 
the cluster, so must detect the emission from the cluster as a whole.  In addition to the known
radio MSP in 47 Tuc, Chandra has detected as many as 20 other X-ray point sources with very soft 
spectra which are thought to also be MSPs.  It is unknown how many undetected MSPs reside in globular
clusters.  Dynamical evolution studies (Ivanova et al. 2004) have estimated that there may be as
many as 200 MSPs in 47 Tuc.  Even though the relatively narrower radio beams may not be visible
in many MSPs, we have argued that all MSPs will have a CR component peaking around 1-10 GeV from 
accelerated primary electrons that is nearly isotropic.  We would then expect the CR emission to be
visible not only from the known radio MSP in clusters, but also from the undetected MSP.  Table 1
lists the known radio MSPs in 47 Tuc with their measured periods and limits on intrinsic period 
derivative derived from a cluster potential model (Freire et al. 2001).  We also list the CR peak
energy and flux above 1 GeV and 10 GeV for each pulsar, where we have used eqs (\ref{ecr}) and 
(\ref{FCR}).  The sum of the flux from these pulsars is estimated, assuming a solid angle of 1 sr. 
The upper limits on emission from 47 Tuc from
EGRET are near our estimates for a solid angle of $2\pi$ sr.  
AGILE and GLAST sensitivity thresholds of $5 \times 10^{-8}\,\rm 
cm^{-2}\,s^{-1}$ and $2 \times 10^{-9}\,\rm cm^{-2}\,s^{-1}$ respectively are well below the 
predicted flux.

\section{Radio-Quiet MSPs}

Many theories of pulsar emission require electron-positron pairs as a necessary ingredient for
coherent radio emission.  As pulsars evolve from left to right in the $P$-$\dot P$ diagram, 
they will move below the ICS pair death line and cease to be radio pulsars.
They will also have no X-ray synchrotron emission components from pairs, nor any synchrotron radiation
component from primary electrons, from resonant absorption of radio emission.
However, these MSPs will still have significant particle acceleration and CR components from the 
primaries.  Although the luminosity will be somewhat lower that the luminosity of MSP above the
death line, some MSPs below the death line have similar spin-down energy to conventional middle-aged
pulsars.  The CR fluxes of some of the nearby X-ray-quiet and radio-quiet MSPs may be detectable by
$\gamma$-ray telescopes with enough sensitivity above 1 GeV.

For example, according to the predictions of equations (\ref{ecr}) and (\ref{FCR}), a MSP pulsar below
the pair death line having period, $P = 3$ ms, $\dot P = 3 \times 10^{-21}\,\rm s\,s^{-1}$ and $B = 2 \times 10^8$
G will have a flux at 100 MeV of $F(100\, {\rm MeV}) = 2 \times 10^{-8}\,\rm ph\,cm^{-2}\,
s^{-1}\,MeV^{-1}\,d_{0.1}^{-2}$ and at 1 GeV of $F(1\, \rm{ GeV}) = 5 \times 10^{-9}\,\rm ph\,cm^{-2}
\,s^{-1}\,MeV^{-1}\,d_{0.1}^{-2}$, where $d_{0.1}$ is the pulsar distance in units of 100 pc. 
A MSP pulsar having period , 
$P = 10$ ms, $\dot P = 1 \times 10^{-20}\,\rm s\,s^{-1}$  and $B = 6 \times 10^8$ G will have a flux at 100 MeV of 
$F(100 \,{\rm MeV}) = 5 \times 10^{-9}\,\rm ph\,cm^{-2}\,s^{-1}\,MeV^{-1}\,d_{0.1}^{-2}$ and at 1 GeV of $F(1\,
 {\rm GeV}) = 1 \times 10^{-9}\,\rm ph\,cm^{-2}\,s^{-1}\,MeV^{-1}\,d_{0.1}^{-2}$.  
Therefore, some radio-quiet MSPs within a 
distance of 100-200 pc will be detectable with GLAST.  Figure 6 shows the expected spectrum of a 
radio-quiet MSP at a distance of 100 pc compared to the GLAST sensitivity limit.  The
sensitivity of ground-based Cherenkov imaging arrays, such as VERITAS, will not reach low enough energy
to detect such sources, but the MAGIC telescope may achieve sensitivity well below 100 GeV.
Surprisingly then, high-energy $\gamma$-ray telescopes will be the only instruments capable of detecting
this hidden population of MSPs.  However, the pulsations will be harder to detect since the number of
photons required for a blind period search is greater than the number required for a point source detection.
However, the number of radio-quiet MSPs below the death line is very uncertain and some may be close
enough to give large photon fluxes.  There may also be MSPs above the death line that are radio-quiet
because the radio beam is small compared to the very broad expected $\gamma$-ray beam.  In this case, the
higher intrinsic luminosities would make these sources easier to detect in pulsed emission and they would
have X-ray SR components.

\section{Conclusions}

We have investigated high-energy emission from MSPs, based on a model for acceleration in the open
field line region above the PC.  The spectral properties of the emission can vary widely,
depending on the position of an individual pulsar relative to the death lines for CR and ICS pairs.
Most MSPs are below the CR pair death line, have unscreened accelerating electric fields and therefore
the primary electrons and pairs accelerate to high altitude above the PC.  The resulting
high-energy spectra in this case consist of several radiation-reaction limited components: CR and ICS from primary
electrons and SR from accelerated primary and secondary electrons.  
The size of the SR component depends on the radio photon luminosity and the multiplicity 
of CR and ICS pairs.  We find in particular that PSR J0218+4232, has a high enough radio luminosity
to form a detectable SR component that matches the observed
X-ray and $\gamma$-ray spectrum.  On the other hand, MSPs such as PSR J0437--4715 that are further
below the CR death line and have a lower radio luminosity have a much weaker pair SR component.  The
X-ray spectra of such MSPs are therefore expected to be dominated by a thermal component from
PC heating.  MSPs such as PSR B1821--24 that are above the CR death line have screened accelerating
fields, so that the spectra are dominated by the radiative energy losses of the particles.

Most of the calculations presented in this paper show the total radiation spectrum and do not necessarily
represent the radiation that is seen by an observer at a particular viewing angle.  Producing such
results will require a full 3D simulation of the pulsar radiation, and we intend to address this more 
realistic calculation in the near future.  Other important relativistic effects such as aberration,
time-of-flight delays, light bending and magnetic-field sweepback will also be significant
(e.g. Dyks \& Harding 2004) and should be included.

The distribution of power in the spectra of MSPs, in which the primary CR spectrum peaks in power 
around 10 GeV, may account for the fact that EGRET 
did not detect many of these sources.  The power gap in MSP spectra falls at 0.1 - 10 GeV, at the
maximum of the EGRET sensitivity range.  The sensitivity of future $\gamma$-ray telescopes will extend
beyond 10 GeV, so that the CR emission from MSPs may become detectable.  The AGILE telescope (Tavani 2003) is
expected to have sensitivity up to 50 GeV and the GLAST LAT sensitivity may reach 300 GeV (McEnery et al. 2004).  
The predicted CR component in MSP spectra has also been out of reach of ground-based Cherenkov telescopes,
that are most sensitive around 1 TeV, and most have poor sensitivity below 100 GeV.   However, several
ground-based telescopes such as MAGIC (Lorenz 2004), STACEE (Williams et al. 2004) and H.E.S.S. (Hinton et al. 2004) 
may have enough sensitivity below 
100 GeV to detect MSPs.  If the CR component from primary electrons is detected in known radio-loud MSPs,
then there is very likely to be a population of radio-quiet MSPs detectable only by high-energy $\gamma$-ray
telescopes, an exciting prospect which can be investigated in the near future.

\acknowledgments 
We would like to thank Yuri Lyubarsky, Roger Romani, Okkie DeJager and Jarek Dyks 
for valuable discussions as well as an anonymous referee for suggesting
critical improvements and corrections to the calculation of the particle pitch angle evolution.  
This work was partially supported by the Israel Science
Foundation of the Israel Academy of Science and Humanities.
We also acknowledge support from the NASA Astrophysics Theory Program.

\appendix
\section{Charged Particle Motion in an Electric and Magnetic Field}

The motion of a charged particle in external electric
(${\bf E}$) and magnetic (${\bf B}$) fields is described by 
the equation

\be  \label{dp}
{d{\bf p}\over dt}=e\{ {\bf E}+ {1\over c}[{\bf v\, B}]\}
+{\bf f}^{\rm rad}\,,
\ee
where ${\bf p}=\gamma m{\bf v}$ is the particle momentum and
${\bf f}^{\rm rad}$ is the radiative reaction force (e.g.,
Landau \& Lifshitz, 1987).

We introduce a right-handed triad of mutually perpendicular 
unit vectors:

\be  \label{triad}
{\bf h}={{\bf B}\over B}\,,\,\,\,\,\,\,{\bf n_1}={{\bf v}_\perp
\over v_\perp}\,\,\,\,\,\,\,{\bf n_2}=[{\bf h\, n_1}]\,.
\ee

In this notation

\be
{\bf p}= p_\parallel{\bf h} +p_\perp {\bf n_1}\,.
\ee

From (\ref{dp}) and (\ref{triad}), we have 

\be  \label{dp_triad}
{dp_\parallel\over dt}{\bf h}+p_\parallel {d{\bf h}\over dt}
+{dp_\perp\over dt}{\bf n_1}+p_\perp{d{\bf n_1}\over dt}=
e\{{\bf E}+{v_\perp\over c}[{\bf n_1}\,{\bf B}]\}
+{\bf f}^{\rm rad}\,.
\ee

Taking the scalar product of equation (\ref{dp_triad}) with ${\bf h}$ and 
then with ${\bf n_1}$  we find

\be  \label{dp_par1}
{d p_\parallel \over dt}
= e ({\bf E\,h}) + p_\perp\left({\bf n_1 \cdot}
{d{\bf h}\over dt}\right)
+({\bf f}^{\rm rad}{\bf \cdot h})\,
\ee 

\be  \label{dp_perp1}
{d p_\perp \over dt}
= e({\bf E\,n_1}) -p_\parallel \left({\bf n_1\cdot}
{d{\bf h}\over dt}\right)
+({\bf f}^{\rm rad}{\bf \cdot n_1})\,, 
\ee
where we use the relation 

\be
{\bf h}{d{\bf n_1}\over dt}= - {\bf n_1}{d{\bf h}\over dt}\,,
\ee
that follows from $({\bf n_1\,h})=0$.

For solution of equations (\ref{dp_par1}) and (\ref{dp_perp1}), 
it is necessary to substitute

\be \label{dh}
{d{\bf h}\over dt}={\partial {\bf h}\over \partial t}
+({\bf v\,\nabla}){\bf h}={\partial {\bf h}\over \partial t}
+v_\parallel ({\bf h\,\nabla}){\bf h} +v_\perp ({\bf n_1\,\nabla})
{\bf h}\,.
\ee

For the application in this paper, we can use the drift approximation 
that is correct with an accuracy of $\sim$ Larmor radius/r.

In this approximation, from equations (\ref{dp_par1}), (\ref{dp_perp1}) 
and (\ref{dh}), the 
following equations for the smoothed values $\bar p_\parallel,\,
\bar p_\perp$ and $v_\perp$ may be derived (e.g., Sivukhin, 1965)

\be  \label{p_par2}
{d p_\parallel\over dt}
=e({\bf E\,h}) +{1\over 2}p_\perp v_\perp {\rm div}\,{\bf h}
+f_\parallel^{\rm rad}\,
\ee

\be  \label{p_perp2}
{d p_\perp \over dt} 
= -{1\over 2}p_\parallel v_\perp {\rm div}\,{\bf h}
+f_\perp^{\rm rad}\,
\ee
where we omit the sign bar, and (Landau \& Lifshitz, 1987)

\be  \label{fpar}
f_\parallel =- {2e^4\over 3m^2c^4}B^2\gamma^2\sin^2 \psi
\cos\psi\,,
\ee

\be  \label{fperp}
f_\perp =- {2e^4\over 3m^2c^4}B^2\gamma^2\sin^3 \psi
\,.
\ee
From equation (\ref{p_perp2}) we can see that $\dot p_\perp$ doesn't 
depend on ${\bf E}$. The field $E_\perp$ enters only into 
the velocity of the guiding center (Sivukhin, 1965). 
Equations (\ref{p_par2}) and (\ref{p_perp2}) coincide with equations (5.4) in
(Sivukhin, 1965). The radiative reaction force is only added
in equations (\ref{p_par2}) and (\ref{p_perp2}).

Since div~${\bf B}=0$, then 

\be
{\rm div}\,{\bf h} = {\rm div} {{\bf B}\over B}=
\left({\bf B\cdot\nabla}{1\over B}\right)
\,.
\ee

The equation (\ref{p_par2}) can be replaced by an equivalent equation that 
is the energy conservation equation. It can be obtained by 
multiplying (\ref{p_par2}) and (\ref{p_perp2}) by $p_\parallel$ and $p_\perp$,
respectively, and adding. This procedure yields

\be  \label{dpsq}
{d\over dt}\left({p^2\over 2}\right)= e p_\parallel ({\bf E\,h})
+f^{\rm rad}p
\ee

or 

\be  \label{gamma2}
{d\over dt}(\gamma mc^2)=e({\bf Ev}_\parallel)+f^{\rm rad}v\,.
\ee
This equation is valid in the drift approximation for
arbitrary electric and magnetic fields. 

Below, we consider motion of ultra relativistic electrons
($\gamma \gg 1$) with very small pitch angles ($\psi \ll 1$)
and their synchrotron emission in the magnetospheres of 
millisecond pulsars. In this case, from equations (\ref{p_perp2}) 
- (\ref{fperp}) and (\ref{gamma2}) we have

\be \label{gamma3}
{d\gamma\over dt}={eE_\parallel\over mc}-{2e^4\over 3m^3c^5}B^2\,p_\perp^2\,, 
\ee

\be  \label{dp_perp3}
{dp_\perp\over dt}=
-{3\over 2}{c\over r}\,p_\perp
-{2e^4\over 3m^3c^5}B^2\,p_\perp^3
{1\over \gamma}\,,  
\ee
To get equation (\ref{dp_perp3}) we have used the relation ${\rm div}\,{\bf h}
\simeq 3/r$ that is valid for the region of particle outflow 
in the pulsar magnetosphere with a dipole magnetic field
at the distance $r$ not too close to the light cylinder.

\newpage
\figureout{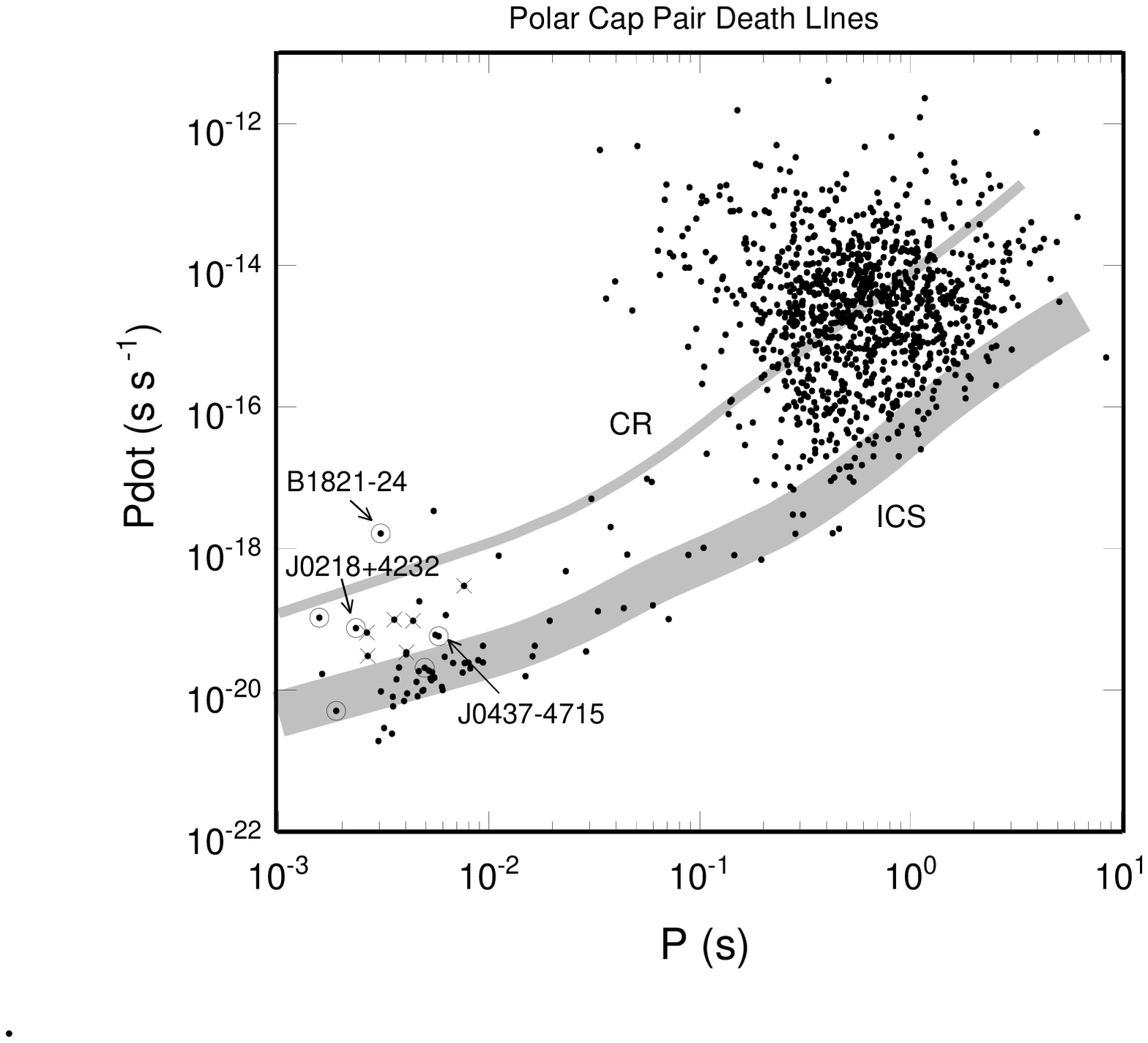}{0}{
Plot of $P$-$\dot P$ for known radio pulsars in the ATNF catalog (http://www.atnf.csiro.au/research/pulsar/psrcat/) 
with measured period derivative.  
Superposed are the pair death lines for curvature radiation (CR) and inverse-Compton scattered (ICS)
photons from Harding et al. (2002).  The width of the death lines indicates the range of uncertainty
due to unknown values of NS surface temperature, mass, radius and moment of inertia.
    }    

\psfig{figure=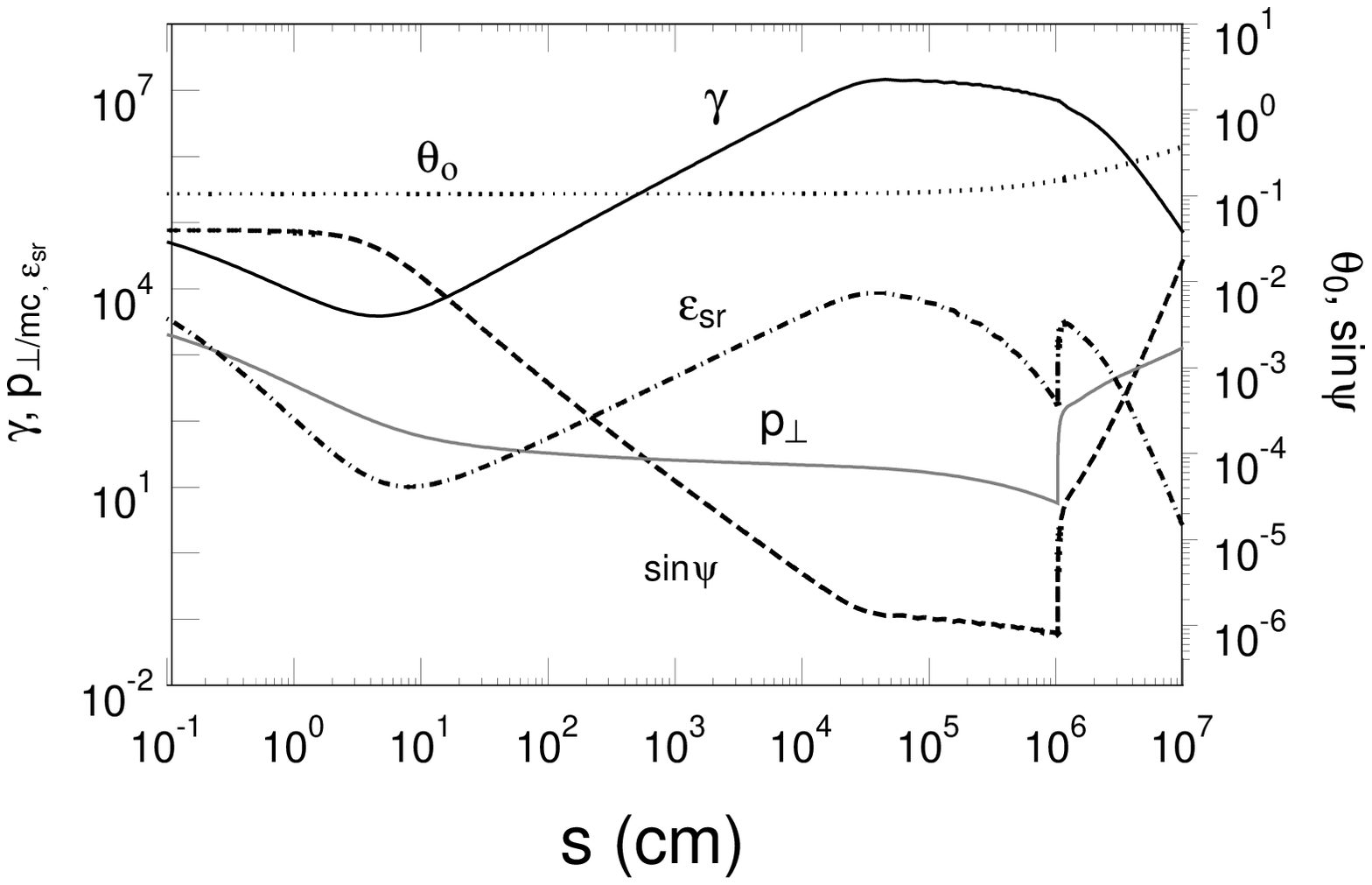,width=5.5in,angle=0}
\figureout{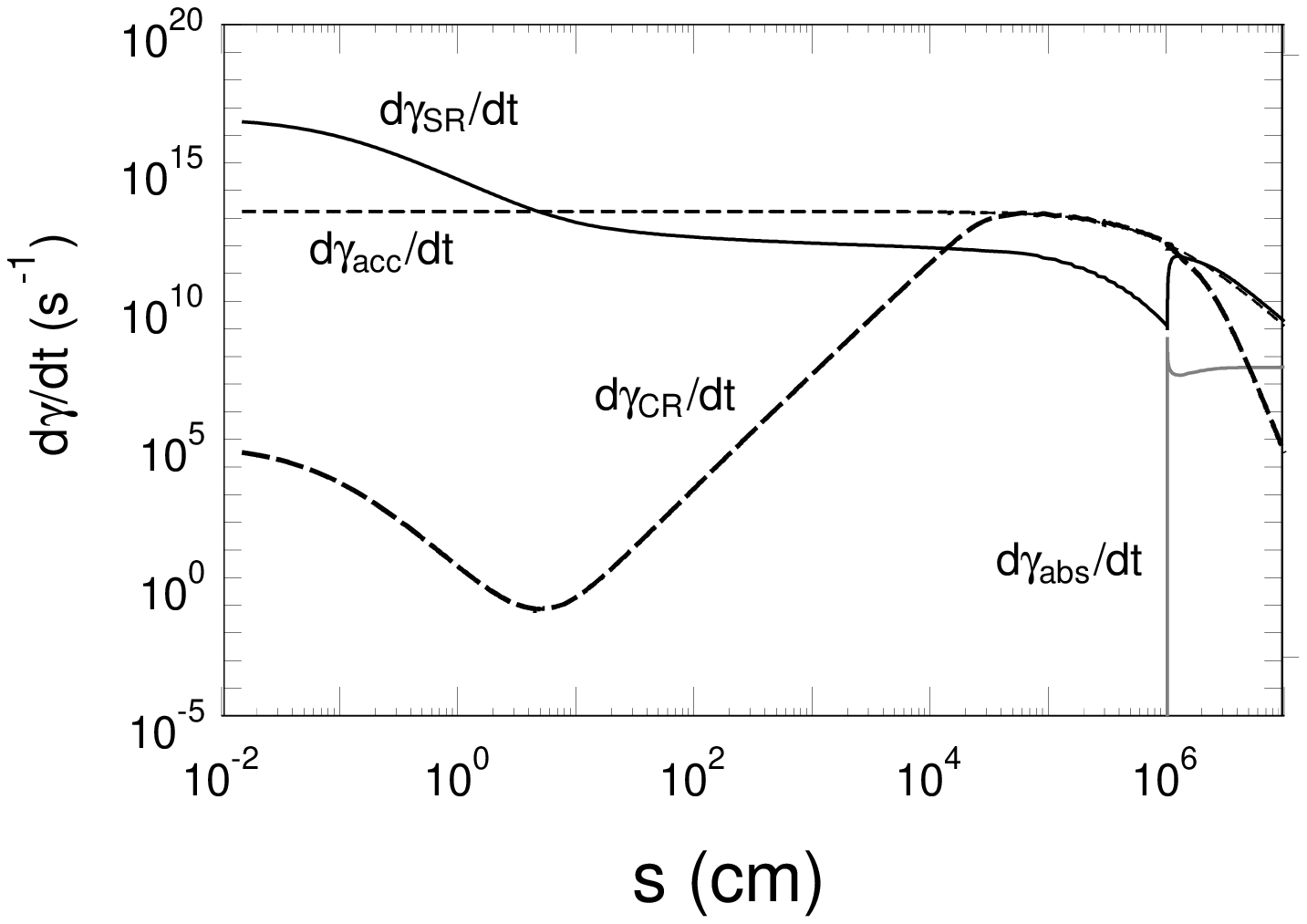}{0}{
Evolution of particle dynamics along a magnetic field line from the NS surface to near the light 
cylinder for the parameters of PSR J0218+4232 and assuming $\nu = 2$. 
a) Lorentz factor $\gamma$ (black solid line), 
momentum perpendicular to the magnetic field 
$p_\perp$ (solid gray line) and pitch angle $\sin\psi$ (black dashed line), the angle between the radio
photons and the particle momentum (dotted line) and the synchrotron radiation critical energy (dot-dashed line) 
as a function of distance along 
the field line.  b) Rate of synchrotron loss (black solid line), curvature radiation loss (thick black dashed
line), acceleration energy gain (thin black dashed line) and cyclotron absorption energy gain (gray solid line).
}    

\psfig{figure=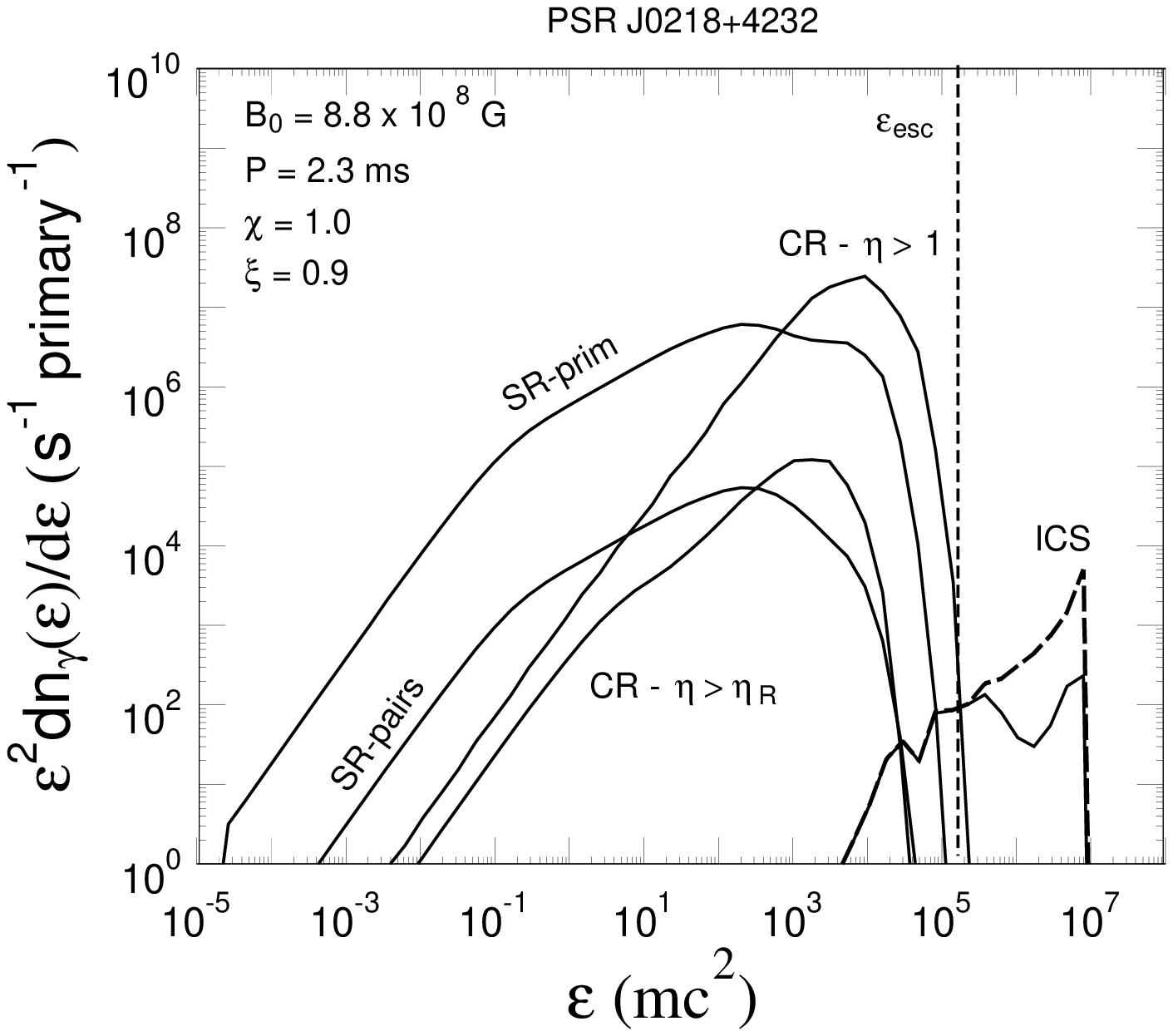,width=5.5in,angle=0}
\figureout{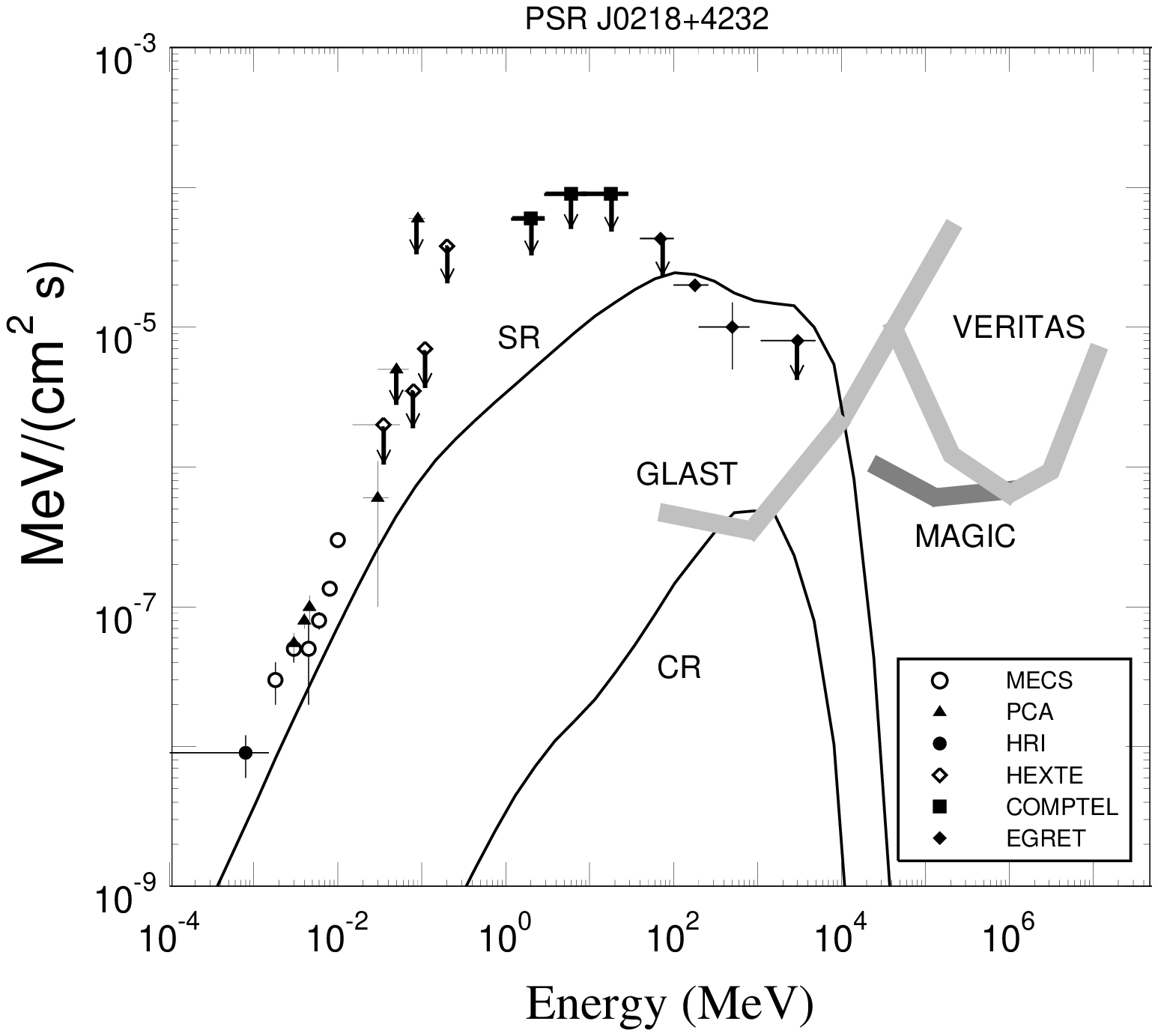}{0}{
a) Radiated model spectrum resulting from unscreened acceleration of a single primary electron, from the
NS surface to the light cylinder, along a field line defined by magnetic colatitude $\xi = 0.9$ in units
of PC half angle.  The pulsar parameters are those of PSR J0218+4232.  The four spectral components shown
are curvature radiation (CR) and inverse-Compton radiation (ICS) from primary electrons (assuming a hot
PC of temperature 1 MK), and synchrotron radiation by primary electrons (SR-prim) and pairs (SR pairs).  
The dashed lines are the unattenuated spectra and the solid lines are the attenuated (escaping) spectra. 
The curvature radiation spectrum labeled CR-$\eta > 1$ is the total spectrum emitted along the field line,
while the spectrum labeled CR-$\eta > \eta_R$ is the spectrum emitted only above the radio emission altitude.    
The escape energy for pair attenuation, $\varepsilon_{\rm esc}$, is indicated by a thin vertical dotted line.
b) Model CR spectrum and pair SR spectrum (solid lines) for inclination 
$\chi \sim 50^{\circ}$ and magnetic colatitude
$\xi = 0.9$, compared to measured spectrum of PSR J0218+4232.  
See text for details on the normalization of the model spectrum.  Data points are from Kuiper et al. (2003).
    }    

\figureout{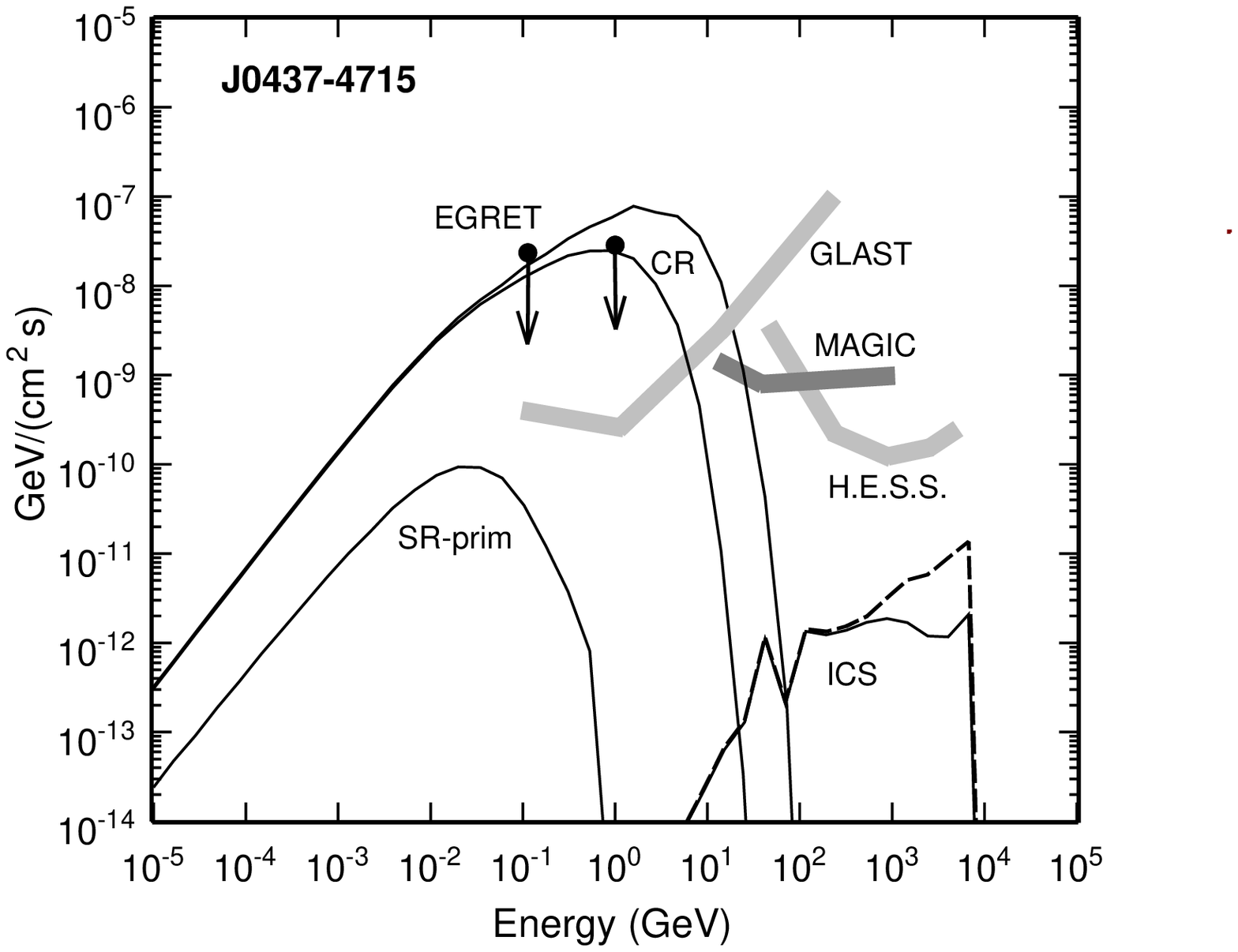}{0}{
Radiated model spectrum for PSR J0437--4715, for unscreened acceleration of a single primary electron,
and for inclination $\chi \sim 10^{\circ}$ and magnetic colatitude of $\xi = 0.5$.
The upper curvature 
radiation (CR) spectrum is the total emission along the field line, while the lower CR spectrum is 
emission only above the radio emission altitude $\eta_R = 2$. 
}    

\figureout{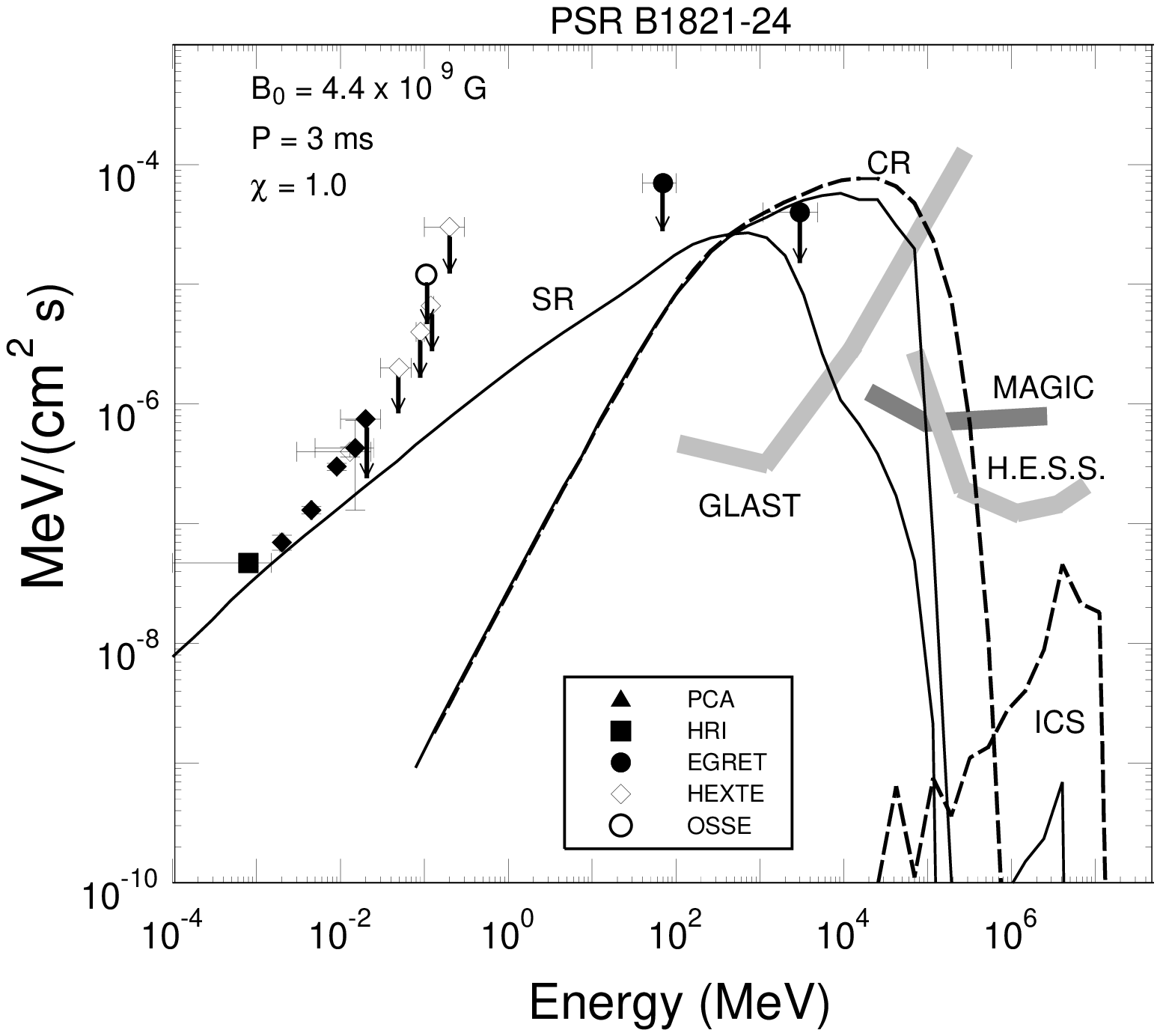}{0}{
Radiated model spectrum for PSR B1821--24, for screened acceleration (i.e. acceleration only near the NS
surface) of a single primary electron, inclination $\chi = 50^{\circ}$ and magnetic colatitude $\xi = 0.7$.
}    

\figureout{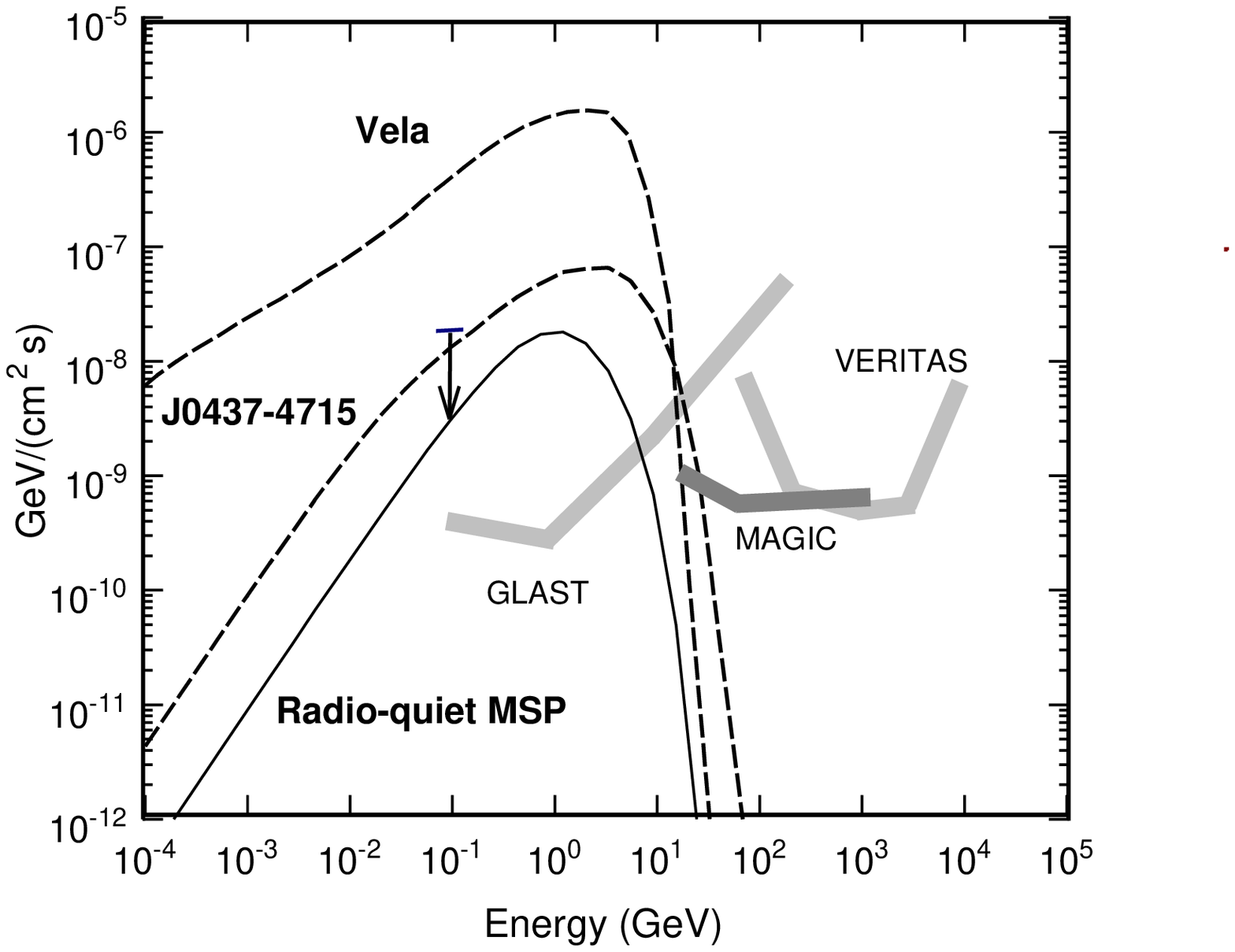}{0}{
Model spectrum for a radio-quiet millisecond pulsar with period $P = 3$ ms and surface magnetic field
$B_0 = 2 \times 10^8$, $\chi = 50^{\circ}$ and magnetic colatitude $\xi = 0.7$.
}    

\newpage
\hoffset -2.0cm
\begin{table}
\caption{Predicted $\gamma$-ray Flux from Millisecond Pulsars}
\begin{tabular}{lcccccccccc}
\hline
PSR	&	$P$ 	&	$\dot P$	&	     d 	&	$B_0$	&	$\varepsilon^{CR}_{\rm peak}$	&  {\small $F_{CR}^*$}	&  $F_{CR}^*$	&  $F_{CR}^*$	& {\small EGRET limit} \\
& (ms) & ($\rm s\,s^{-1})$ & (kpc) & ($10^8$ G) & (GeV) &  ({\small 100 MeV}) & (1 GeV) & ($>${\small 100MeV}) & 	($>$ {\small 100 MeV}) \\
	&		&		&	 	&  		& 		&  \multicolumn{2}{c}{{\small $10^{-8}\,\rm ph/(cm^2\,s\,MeV$))}}	&  \multicolumn{2}{c}{{\small ($10^{-8}\,\rm ph/(cm^2\,s$))}}		\\
\hline
Plane	&		&		&		&		&		&		&		&		&		\\	
J0437-4715	&	5.75	&	0.205	&	0.18	&	6.95	&	4.81	&	0.672	&	0.145	&	531.573	&	21	\\	
J1821-24	&	3.05	&	16	&	5.5	&	44.7	&	42.90	&	0.020	&	0.004	&	39.918	&	58	\\	
J0218-4232	&	2.32	&	0.75	&	5.85	&	8.0	&	17.30	&	0.006	&	0.001	&	8.816	&		\\	
B1937+21	&	1.6	&	1	&	3.60	&	8.10	&	26.67	&	0.039	&	0.008	&	63.176	&	58	\\	
B0030+0451	&	4.86	&	0.1	&	0.23	&	4.46	&	4.25	&	0.391	&	0.084	&	292.484	&		\\	
J2124-3358	&	4.93	&	1.077	&	0.25	&	1.47	&	10.24	&	1.059	&	0.228	&	1168.888	&		\\	
B1957+20	&	1.607	&	0.1685&	1.6	&	3.33	&	13.63	&	0.080	&	0.017	&	99.311	&		\\	
	&		&		&	 	&	 	&	 	&	 	&	 	&	 	&		\\	
47 Tuc	&		&	 	&		&		&	 	&	 	&	 	&	 	&		\\	
J0023-7204C	&	5.757	&	0.24	&	5	&	7.52	&	5.09	&	0.001	&	0.000	&	0.764	&		\\	
J0023-7204D	&	5.358	&	1.3	&	5	&	16.9	&	10.22	&	0.002	&	0.001	&	2.753	&		\\	
J0023-7204E	&	3.536	&	1.9	&	5	&	16.6	&	16.95	&	0.006	&	0.001	&	8.798	&		\\	
J0023-7204F	&	2.624	&	1.9	&	5	&	14.3	&	22.01	&	0.011	&	0.002	&	16.888	&		\\	
J0023-7204G	&	4.04	&	1.3	&	5	&	14.7	&	13.09	&	0.004	&	0.001	&	5.125	&		\\	
J0023-7204H	&	3.21	&	0.7	&	5	&	9.59	&	12.69	&	0.005	&	0.001	&	5.660	&		\\	
J0023-7204I	&	3.485	&	1	&	5	&	11.9	&	13.50	&	0.005	&	0.001	&	5.970	&		\\	
J0023-7204J	&	2.101	&	0.24	&	5	&	4.54	&	12.31	&	0.006	&	0.001	&	7.117	&		\\	
J0023-7204L	&	4.346	&	1	&	5	&	13.3	&	11.13	&	0.003	&	0.001	&	3.674	&		\\	
J0023-7204M	&	3.677	&	0.18	&	5	&	5.21	&	6.77	&	0.002	&	0.000	&	1.710	&		\\	
J0023-7204N	&	3.054	&	0.8	&	5	&	10.0	&	13.93	&	0.005	&	0.001	&	6.892	&		\\	
J0023-7204O	&	2.643	&	1.7	&	5	&	13.6	&	20.98	&	0.010	&	0.002	&	15.465	&		\\	
J0023-7204Q	&	4.033	&	1	&	5	&	12.9	&	11.88	&	0.004	&	0.001	&	4.331	&		\\	
J0023-7204T	&	7.588	&	6	&	5	&	43.2	&	13.38	&	0.003	&	0.001	&	3.498	&		\\	
J0023-7204U	&	4.343	&	1.7	&	5	&	17.4	&	13.58	&	0.004	&	0.001	&	5.213	&		\\	
	&		&		&		&		&		&		&		&		&		\\	
SUM	&		&		&		&		&		&	0.072	&	0.016	&	93.856	&	5	\\	
\hline
\end{tabular}
$^*$ Assuming a 1 sr. solid angle and $\eta_{\rm max} = 2$

\end{table}

\end{document}